\documentclass[preprint,aps,prd,groupedaddress,showpacs,nofootinbib]{revtex4-1}
\usepackage{hyperref}
\usepackage{pifont}
\usepackage{amsmath,amsfonts,amssymb,amsthm,mathrsfs}
\usepackage{graphicx,chngcntr,vmargin,braket}
\usepackage[usenames]{xcolor}
\setmarginsrb{20mm}{10mm}{20mm}{20mm}{10mm}{10mm}{10mm}{10mm}
\allowdisplaybreaks[1]

\newcommand{\oo}{\infty}
\renewcommand{\d}{\mathrm{d}}
\newcommand{\bea}{\begin{eqnarray}}
\newcommand{\eea}{\end{eqnarray}}

\newcommand{\Tauc}{\mathcal{T}_{\text{cut}}}
\newcommand{\Tau}{\mathcal{T}}

\makeatletter
\newcommand\footnoteref[1]{\protected@xdef\@thefnmark{\ref{#1}}\@footnotemark}
\makeatother

\begin{document}

\title{Next-to-leading power corrections to $V+1$ jet production in
  $N$-jettiness subtraction}
\author{Radja Boughezal}
\affiliation{High Energy Physics Division, Argonne National Laboratory, Argonne, IL 60439, USA}
\author{Andrea Isgr\`o}
\affiliation{Department of Physics \& Astronomy, Northwestern University, Evanston, IL 60208, USA} 
\author{Frank Petriello}
\affiliation{High Energy Physics Division, Argonne National Laboratory, Argonne, IL 60439, USA}
\affiliation{Department of Physics \& Astronomy, Northwestern University, Evanston, IL 60208, USA}

\begin{abstract}

  We discuss the subleading power corrections to one-jet production
  processes in $N$-jettiness subtraction using vector-boson plus jet
  production as an example.  We analytically derive the
  next-to-leading power leading logarithmic corrections (NLP-LL) through ${\cal
    O}(\alpha_S)$ in perturbative QCD, and outline the calculation of the next-to-leading logarithmic corrections (NLP-NLL). 
  Our result is differential in the jet transverse momentum and rapidity, and
  in the vector boson momentum squared and rapidity.
  We present simple formulae
  that separate the NLP corrections into universal factors valid for
  any one-jet cross section and process-dependent matrix-element
  corrections.  We discuss in detail features of the NLP corrections such as 
  the process independence of the leading-logarithmic result that
  occurs due to the factorization of matrix elements in the subleading
  soft limit,
  the occurrence of poles in the non-hemisphere soft function at NLP and
   the cancellation  of potential $\sqrt{\Tau_1/Q}$
 corrections to the $N$-jettiness factorization theorem. We validate
 our analytic result by comparing them to numerically-fitted 
 coefficients, finding good agreement for both the inclusive and the
 differential cross sections.  The size of the power
   corrections for different definitions of $\Tau_1$ is studied.

\end{abstract}

\date{\today}
\maketitle
\tableofcontents

\section{Introduction}

There has been significant recent interest in the study of
subleading power corrections to factorization theorems in QCD.  This
focus is driven in large part by the increasingly precise data
delivered by the Large Hadron Collider (LHC).  Obtaining theoretical
predictions that match the experimental precision increasingly
requires going beyond the leading-power formalisms that underly past
theoretical calculations.  One recent example is the study of
next-to-leading power corrections to the $N$-jettiness factorization
theorem~\cite{Moult:2016fqy,Boughezal:2016zws,Moult:2017rpl,Feige:2017zci,Moult:2017jsg,Boughezal:2018mvf,Ebert:2018lzn}
that underlies the $N$-jettiness subtraction method for precision cross
section calculations~\cite{Boughezal:2015dva,Gaunt:2015pea}.  Other
results include initial studies of the subleading power corrections
to the low transverse momentum factorization
theorem~\cite{Ebert:2018gsn,Cieri:2019tfv}, and the study of subleading power
corrections to threshold production of color-singlet states~\cite{Bonocore:2014wua,Bonocore:2015esa,Bonocore:2016awd,DelDuca:2017twk,Beneke:2018gvs}.

A feature of these improvements is that they are limited to
color-singlet processes without jets in the final state.  Relatively few
results for subleading corrections to jet production processes are
available, although some studies of jet
production at subleading power have recently been
initiated~\cite{Beneke:2017ztn,Beneke:2018rbh,vanBeekveld:2019prq}.  A understanding of the
next-to-leading power corrections to the $N$-jettiness factorization
theorem~\cite{Stewart:2009yx,Stewart:2010tn} in the presence of
final-state jets is highly desirable.  Although $N$-jettiness
subtraction has been used to derive the next-to-next-to-leading order
perturbative QCD corrections needed to properly describe hadron
collider data for a host of processes~\cite{Boughezal:2015dva,Boughezal:2015aha,Boughezal:2015ded,Boughezal:2016yfp,Boughezal:2016dtm,Boughezal:2016isb,Abelof:2016pby,Campbell:2016lzl,Boughezal:2017nla,Campbell:2018wfu,Campbell:2019gmd}, these applications are
computationally intensive.  One approach
to improve computational efficiency is to analytically calculate the
power corrections.  This extends the region of validity of the
factorization theorem to higher $N$-jettiness values, ameliorating the
difficulties that arise from numerically extracting the large
logarithms of $N$-jettiness that appear in the individual cross
section components.

In this paper we take a first step towards understanding the
subleading power corrections to jet production processes in
$N$-jettiness subtraction by computing the next-to-leading power (NLP) corrections to the one-jettiness
factorization theorm at next-to-leading order (NLO) in perturbative
QCD.  Our primary results are simple analytic formulae for the
leading-logarithmic (LL) power corrections.  We additionally outline the extension
of this calculation to NLL.  We separate the power corrections into
process-independent terms valid for any one-jet production process and
process-dependent matrix element correction factors.  Important aspects of
our results are summarized below.
\begin{itemize}

 \item We make use of the expansion by regions~\cite{Beneke:1997zp,Jantzen:2011nz}
 to perform the computation of the cross section. In particular, we split the phase
 space into two beam regions, a jet region and a soft region.

  \item We show that all NLP-LL corrections at NLO arise from the
    emission of soft partons, as in the case of color-singlet
    production~\cite{Moult:2016fqy, Boughezal:2016zws}, and show how to obtain such subleading soft corrections by making use of
  the subleading soft theorem~\cite{Burnett:1967km}.  This allows us
  to write the NLP-LL result in a universal form valid for all one-jet processes.

 \item We show that the non-hemisphere soft contributions defined
   in~\cite{Jouttenus:2011wh}, which are finite at leading power, contribute to poles
   when extended to next-to-leading power.  These poles are necessary
   for the consistency of the result at NLP.
  
  \item We demonstrate the cancellation of potential power corrections
    suppressed only by $\sqrt{\Tau/Q}$, where $\Tau$ is the
    one-jettiness event shape variable and $Q$ is a generic hard
    scale.

\end{itemize}  
Our paper is organized as follows.  In Section~\ref{sec:Born} we
discuss the Born-level process for $V+j$ production and introduce the
notation used in the remainder of the manuscript.  We discuss our
strategy for the computation of the NLP corrections in
Section~\ref{sec:strategy}, and illustrate the separation of the phase space
into different regions. In Section~\ref{sec:PS}, we write down
a general expression for the phase space that is valid in every region,
separating the case where the two final-state partons are measured as
two separate jets from the case where they are part of the same jet.
We then proceed to expand the phase space  in each region, listing all the relevant expansion
coefficients in the Appendix.  We discuss the expansion of the matrix elements in
Section~\ref{sec:ME}, providing the explicit expression of all the relevant expansion coefficients in a supplemental file.  An important aspect of this section is
how the soft expansion can be predicted by the subleading
soft theorem, without needing the full NLO amplitude.  This leads to a
simple, universal expression for the NLP-LL result.  
There is currently no subleading collinear factorization
theorem for QCD amplitudes, which is required for a similar universal
description of the NLP-NLL result.  The beam and jet expansions to the
NLP-NLL level
therefore require us to use the full NLO amplitude.
In Section~\ref{sec:LP} we derive as a check on our result the
leading-power cross section.  The primary results
of our paper, which are the analytic forms of the NLP-LL corrections,
are presented in Section~\ref{sec:NLP}.  In Section~\ref{sec:numerics}
we provide numerical checks of our analytic results, for both the inclusive and the
differential cross section.
Finally, we conclude in Section~\ref{sec:conc}.

\section{Description of the Born-level process}
\label{sec:Born}

We will illustrate our derivation of the NLP corrections using 
$V+j$ production as an example. We show formulae for the Born-level partonic
process is $q \left( q_a\right) + \bar{q} \left( q_b\right) \to V
\left( p_V\right) + g\left( q_J\right)$, where $V$ is a vector
boson.  The derivation for the quark-gluon Born-level process is identical. We parametrize the momenta in the lab frame of reference:
\begin{equation}
q_a^\mu = \frac{x_a \sqrt{s}}{2} n_a^\mu, \qquad \qquad q_b^\mu = \frac{x_b \sqrt{s}}{2} n_b^\mu, \qquad \qquad q_J^\mu =p_T \cosh \eta \, n_J^\mu, \qquad \qquad p_V^\mu = q_a^\mu + q_b^\mu - q_J^\mu,
\label{eq:bornmomenta}
\end{equation}
where $x_a$ and $x_b$ are the Born momentum fractions of the two
initial-state partons, $\sqrt{s}$ is the energy of the hadronic collision, $p_T$ is the jet transverse momentum and $\eta$ is the jet pseudorapidity.
We have defined the following light-like vectors that describe the two
beam directions and the jet direction:
\begin{equation}
n_a^\mu = \begin{pmatrix} 1\\0\\0\\1\end{pmatrix}, \qquad \qquad
n_b^\mu = \begin{pmatrix} 1\\0\\0\\-1\end{pmatrix}, \qquad \qquad
n_J^\mu = \begin{pmatrix} 1 \\ \frac{1}{\cosh \eta} \\ 0 \\ \tanh \eta \end{pmatrix}.
\label{eq:nmomenta}
\end{equation}
The phase space, including the flux factor and parton distribution
functions (PDFs), takes the form
\begin{align}
\text{PS}_\text{Born} = \frac{1}{(2 \pi)^{d-2}}&\int_0^1 \d x_a \int_0^1 \d x_b \frac{f_q(x_a)f_{\bar{q}}(x_b)}{2 s x_a x_b} \int \d^d p_V \, \delta\left(p_V^2 - m_V^2\right) \nonumber \\
&\int \d^d q_J \, \delta\left(q_J^2\right)  \, M_J \left( q_J\right) \delta^{(d)}(q_a+q_b-q_J-p_V).
\end{align}
Here, $q$ and $\bar{q}$ are the initial-state quark flavors and $M_J \left(
  q_J\right)$ is a jet measurement function that ensures us that $q_J$
is indeed a jet (it can simply be a set of experimental cuts on the
jet $p_T$ and pseudorapidity). In simplifying the Born phase space, we
wish to be differential in the vector boson momentum squared and
rapidity. Those quantities are defined as
\begin{align}
Q^2 &= p_V\cdot p_V=s x_a x_b - \sqrt{s} p_T x_a e^{-\eta}- \sqrt{s} p_T x_b e^{\eta}, \\
Y&= \frac12 \log \left( \frac{p_V^0 + p_V^z}{p_V^0- p_V^z}\right) =\frac12 \log \left(\frac{\sqrt{s} x_a - p_T e^\eta}{\sqrt{s} x_b - p_T e^{-\eta}} \right).
\end{align}
We can use the constraints on these quantities to solve for the Born
momentum fractions of the two initial-state partons:
\begin{align}
x_a &= \frac{p_T e^\eta}{\sqrt{s}} + \frac{e^Y}{\sqrt{s}} \sqrt{p_T^2 + Q^2}, \\
x_b &= \frac{p_T e^{-\eta}}{\sqrt{s}} + \frac{e^{-Y}}{\sqrt{s}} \sqrt{p_T^2 + Q^2}.
\end{align}
Imposing the on-shellness of the final-state gluon, we obtain the
following differential Born-level phase space:
\begin{equation}
\frac{\d \text{PS}_\text{Born}}{\d Q^2 \, \d Y \, \d p_T \, \d \eta} =\frac{\Omega_{d-2} p_T^{d-3}}{2 s (2\pi)^{d-2} }\frac{f_q(x_a) f_{\bar{q}}(x_b)}{2 s x_a x_b} \delta \left(Q^2 - m_V^2\right) M_J \left( p_T, \eta \right).
\end{equation}
For future notational convenience we define a partonic Born-level
phase space with the PDFs removed:
\begin{equation}
\frac{\d \hat{\text{PS}}_\text{Born}}{\d Q^2 \, \d Y \, \d p_T \, \d \eta} \equiv \frac{\Omega_{d-2} p_T^{d-3}}{2 s (2\pi)^{d-2} } \frac{\delta \left(Q^2 - m_V^2\right)}{2 s x_a x_b} M_J \left( p_T, \eta \right).
\end{equation}

The matrix element is a function of the
invariants $s_{ij}$, where $s_{ij}\equiv \left(p_i+p_j \right)^2$ if
both $p_i$ and $p_j$ are initial or final-state partons, and
$s_{ij}\equiv \left(p_i-p_j \right)^2$ if one is an initial-state
parton and the other is a final-state parton. The Born amplitude squared is
\begin{equation}
\mathcal{M}_\text{Born} = N_\text{W} \left(2 C_F\right) \, \frac{2 s_{12}^2 + 2 s_{12} \left(s_{13} + s_{23} \right) + s_{13}^2 + s_{23}^2}{s_{13} s_{23}}
\end{equation}
where $N_\text{W}$ is an electroweak normalization factor. 

\section{Strategy for the computation}
\label{sec:strategy}

At NLO in QCD perturbation theory, the power corrections arise from
real-emission corrections.  To study the structure of the NLP
corrections we consider the real-emisson process
$q(q'_a) + \bar{q}(q'_b) \to V(p_V) + g(p_3) + g(p_4)$ as an example, where the
initial-state momenta are labeled with a prime in order to
distinguish them from the Born initial-state
momenta. The study of the $qg \to Vqg$ process
  proceeds similarly, and we do not present it explicitly.  We will
  later present results valid for both channels.  The one-jettiness event-shape variable $\Tau_1$ can be defined as~\cite{Jouttenus:2011wh}
\begin{align}
\Tau_1 &= \sum_k \min_i \left\{ \frac{2 q_i \cdot p_k}{Q_i}\right\} = \sum_k \min_i \left\{ \frac{n_i \cdot p_k}{\rho_i}\right\}\nonumber \\
&=\min \left\{ \frac{n_a \cdot p_3}{\rho_a},\frac{n_b \cdot p_3}{\rho_b},\frac{n_J \cdot p_3}{\rho_J}\right\} + \min\left\{\frac{n_a \cdot p_4}{\rho_a},\frac{n_b \cdot p_4}{\rho_b},\frac{n_J \cdot p_4}{\rho_J} \right\},
\label{eq:taudef}
\end{align}
where $q_i$ are the two beam momenta and the jet momentum at Born level,
$p_k$ are the final-state parton momenta and $Q_i$ are normalization
factors. We have substituted $2 q_i^\mu/Q_i$ with $n_i^\mu/\rho_i$ for
notational convenience. The light-like momenta $n_a^\mu$,
$n_b^\mu$ and $n_J^\mu$ are the same as in Eq.~\eqref{eq:nmomenta}. From now on, the subscript in $\Tau_1$ will be implicit and we will simply refer to the 1-jettiness as $\Tau$. The measurement of $\Tau$ is encoded in the measurement function
$ \delta \left[\Tau - \hat{\Tau} \left( p_3,p_4\right)\right] $, which,
due to the presence of a jet in the final state, is considerably more involved than in the
0-jettiness case. 

The first simplification to the measurement function comes from
exploiting the symmetry $p_3 \leftrightarrow p_4$ relevant for the
partonic process under consideration. The gluons in the final state are
identical, leading to an overall factor of 1/2 in the cross
section. We can always assume that $p_{3T} \ge p_{4T}$, modulo
relabelling $p_3 \leftrightarrow p_4$. The relabelling freedom cancels
the 1/2 symmetry factor.  The momentum $p_4 \equiv k$ can therefore always be
considered as the emitted gluon which can become soft or collinear,
while $p_3$ can always be considered as a hard parton which is either
the jet itself or its hardest partonic component.

A procedure is needed to determine the jet
momentum at NLO. A clustering algorithm normally defines a distance 
measure between the final state particles. If this distance
is larger than a certain value (e.g. the size of the jet cone) then the two
final-state partons will be clustered as two separate jets, and the parton with
the largest transverse momentum ($p_3$) will be the leading jet. Otherwise, if the distance between
the final-state partons is small, the jet momentum will be the sum of the momenta 
of the two partons.  We find
it simplest to use $N$-jettiness itself as a jet algorithm. The scalar product
$n_J \cdot k$ is indeed a measure of the distance between the two final-state partons.
When this distance is smaller than all the other scalar products that appear in
the one-jettiness definition of Eq.~\eqref{eq:taudef}, then the two final-state partons
are clustered as a single jet whose momentum is $p_3+p_4$. Otherwise, the
two final-state partons form two separate jets.

If the final-state partons are
clustered as separate jets, then $q_J^\mu \equiv p_3^\mu$, since $p_3$
is hard and must therefore be the only jet in the low-$\Tau$
limit. This means that the first minimum in Eq.~\eqref{eq:taudef} is zero,
since $n_J \cdot p_3 = n_3 \cdot p_3= 0$. If the two final-state
partons are instead clustered together in the same jet, then the first
minimum in Eq.~\eqref{eq:taudef} must be $\frac{n_J \cdot p_3}{\rho_J}$,
since $p_3$ is not allowed to be soft or collinear to the beam
direction due to the jet measurement function.  It can, however, be
collinear to the jet direction $n_J$.

These assumptions being made, the measurement function can be written as
\begin{align}
\delta \left[ \Tau - \hat{\Tau} \left( p_3, p_4\right)\right] &= \Theta \left( \Tau_b-\Tau_a\right)\Theta \left( \Tau_J-\Tau_a\right) \delta \left(\Tau - \Tau_a \right) \nonumber \\
&+\Theta \left( \Tau_a-\Tau_b\right)\Theta \left( \Tau_J-\Tau_b\right) \delta \left(\Tau - \Tau_b \right) \nonumber \\
&+\Theta \left( \Tau_a-\Tau_J\right)\Theta \left( \Tau_b-\Tau_J\right) \delta \left(\Tau - \Tau'_J \right) 
\label{eq:measurement}
\end{align}
where we have defined
\begin{equation}
\Tau_i \equiv \frac{n_i \cdot k}{\rho_i}, \qquad \qquad \Tau'_i \equiv \frac{n_i \cdot k}{\rho_i} + \frac{n_J \cdot p_3}{\rho_J}.
\label{eq:tauidef}
\end{equation}
In order to further simplify the measurement function, we will make
use of the expansion by
regions~\cite{Beneke:1997zp,Jantzen:2011nz}. The necessary regions are
listed below.
\begin{itemize}
\item Beam $a$ region: $\Tau_a \ll \Tau_b, \Tau_J$. The measurement function becomes
\begin{equation}
\delta \left[ \Tau - \hat{\Tau} \left( p_3, p_4\right)\right] \to  \delta \left(\Tau - \Tau_a \right).
\end{equation}
\item Beam $b$ region: $\Tau_b \ll \Tau_a, \Tau_J$. The measurement function becomes
\begin{equation}
\delta \left[ \Tau - \hat{\Tau} \left( p_3, p_4\right)\right] \to  \delta \left(\Tau - \Tau_b \right).
\end{equation}
\item  Jet region: $\Tau_J \ll \Tau_a, \Tau_b$. The measurement function becomes
\begin{equation}
\delta \left[ \Tau - \hat{\Tau} \left( p_3, p_4\right)\right] \to  \delta \left(\Tau - \Tau'_J \right).
\end{equation}
\item Soft region: $\Tau_a \sim \Tau_b \sim \Tau_J \ll Q$. The measurement function cannot be expanded 
since all of the terms that appear in it are homogeneous. We make use of the hemisphere
decomposition~\cite{Jouttenus:2011wh} and write the measurement function as
\begin{align}
\delta \left[\Tau - \hat{\Tau} \left( p_3, p_4\right)  \right] &=\Theta \left(\Tau_j-\Tau_i \right)\delta \left(\Tau-\Tau_i \right)\nonumber \\
&+ \Theta \left( \Tau_j - \Tau_i\right) \Theta \left( \Tau_i-\Tau_m\right) \left[\delta \left(\Tau-\Tau_m \right) -\delta \left(\Tau-\Tau_i \right) \right]+ \left(i \leftrightarrow j \right)
\label{eq:measurementhemi}
\end{align}
where with a slight abuse of notation $\delta\left(\Tau-\Tau_J \right)$ is always substituted with $\delta \left( \Tau-\Tau'_J\right)$.
We emphasize that Eq.~\eqref{eq:measurementhemi} is not an expansion,
as for any choice of $i$ and $j$, we reproduce exactly the complete
measurement function Eq.~\eqref{eq:measurement}. The choice of $i$ and $j$ will be different for each term in the integrand, and in Sections~\ref{sec:softLP} and~\ref{sec:NLPsoft} we illustrate how we make this choice.
\end{itemize}
In our computaton we proceed by expanding the phase space and the matrix
element in each region.

\section{NLO phase space}
\label{sec:PS}
The NLO phase space differential in the vector boson momentum squared and rapidity, and in the jet transverse momentum and pseudorapidity, is
\begin{align}
\frac{\d\text{PS}}{\d Q^2 \, \d Y\, \d p_T\, \d \eta} =& \frac{\mu_0^{2\varepsilon}}{(2\pi)^{2d-3}}\int \d \xi_a \int \d \xi_b \frac{f_q(\xi_a)\, f_{\bar{q}}(\xi_b)}{2 s \xi_a \xi_b} \int  \d^d p_V \delta \left(p_V^2 - m_V^2 \right)\int  \d^d p_3  \, \delta \left(p_3^2\right) \nonumber \\
&  \int \d^d p_4 \, \delta \left(p_4^2\right) \delta^{(d)} \left(q'_a +q'_b - p_V - p_3 - p_4 \right) \, M_J \left(q_J \right) \delta \left[\Tau - \hat{\Tau} \left( p_3, p_4\right)  \right]\nonumber \\
& \delta \left( p_V^2 - Q^2\right) \delta \left[Y-\frac12\log \left( \frac{p_V \cdot n_b}{p_V \cdot n_a}\right) \right] \delta \left[ p_T -\hat{p}_T \left( p_3,p_4\right)\right] \delta \left[\eta - \hat{\eta} (p_3,p_4) \right],
\label{eq:NLOPS}
\end{align}
where $\mu_0$ is the minimal subtraction renormalization scale.  The initial-state momentum fractions have been labeled as $\xi_a$, $\xi_b$ in order to distinguish them from the Born initial-state momentum fractions $x_a$, $x_b$. We parametrize  $p_4 \equiv k$ according to a Sudakov decomposition, where the two light-like vectors that describe the directions of the decomposition are in general $n_i^\mu$ and $n_j^\mu$:
\begin{equation}
k^\mu = \frac{k_j}{n_i \cdot n_j} n_i^\mu +\frac{k_i}{n_i \cdot n_j} n_j^\mu + k_\perp^\mu,
\label{eq:Sudakovk}
\end{equation}
The integral in the final-state gluon momentum can then be written as
\begin{equation}
\int \d^d k \, \delta\left(k^2\right) =\frac{\Omega_{d-3}}{4} \left(\hat{s}_{ij} \right)^{-1+\varepsilon}\int_0^\pi \d \phi \left( \sin^2 \phi \right)^{-\varepsilon} \int \d \Tau_i \, \Tau_i^{-\varepsilon} \int \d \Tau_j \, \Tau_j^{-\varepsilon}.
\end{equation}
where we have defined the hatted invariants $\hat{s}_{ij}$ as~\cite{Jouttenus:2011wh}:
\begin{equation}
\hat{s}_{ij} = \frac{n_i \cdot n_j}{2 \rho_i \rho_j}.
\end{equation}

The operators $\hat{p}_T(p_3,p_4)$ and $\hat{\eta}(p_3,p_4)$ in Eq.~\eqref{eq:NLOPS} measure the jet transverse momentum and rapidity. When the two final-state partons are separate, then they are simply $p_T = p_{3T}$ and $\eta = \eta_3$. When the two final-state partons are part of the same jet, we define the jet momentum $q'_J=p_3+p_4$ and then measure its transverse momentum and pseudorapidity. 

In the two-jet case, we change variables from $\xi_a$, $\xi_b$ to
$Q^2$, $Y$, similar to the procedure followed at Born level. When $ij=ab$, for example, this change of variables is
\begin{equation}
\xi_a = \frac{k_b + e^\eta p_T}{\sqrt{s}} + \frac{e^Y}{\sqrt{s}} \sqrt{k_a k_b + p_T^2 + Q^2 + 2 \sqrt{k_a k_b} p_T \cos \phi},
\label{eq:xia}
\end{equation}
\begin{equation}
\xi_b = \frac{k_a + e^{-\eta} p_T}{\sqrt{s}} + \frac{e^{-Y}}{\sqrt{s}} \sqrt{k_a k_b + p_T^2 + Q^2 + 2 \sqrt{k_a k_b} p_T \cos \phi}.
\label{eq:xib}
\end{equation}
The phase space in the two-jet case is then
\begin{align}
\frac{\d\text{PS}_{ij,2J}}{\d Q^2\, \d Y\, \d p_T\, \d \eta} =& \frac{\d\hat{\text{PS}}_\text{Born}}{\d Q^2\, \d Y\, \d p_T\, \d \eta} \frac{\Omega_{d-3}\mu_0^{2\varepsilon}}{4(2\pi)^{d-1}} \left(\hat{s}_{ij} \right)^{-1+\varepsilon}\int_0^\pi \d \phi \left( \sin^2 \phi \right)^{-\varepsilon}\nonumber \\
&  \int \d \Tau_i \, \Tau_i^{-\varepsilon} \int \d \Tau_j \, \Tau_j^{-\varepsilon}\, \, \frac{x_a x_b}{ \xi_a \xi_b}f_i(\xi_a)\, f_j(\xi_b) \delta \left[\Tau - \hat{\Tau}\left( p_3,p_4\right) \right].
\label{eq:PS2jetgeneral}
\end{align}

In the one-jet case, the jet momentum is ${q'_J}^\mu =
p_3^\mu + p_4^\mu$. To derive a convenient form of the phase space we
first parametrize the jet momentum in terms of its transverse mass, transverse momentum and pseudorapidity:
\begin{equation}
{q'_J}^\mu = \begin{pmatrix} m_T \cosh \eta \\ p_T \\ 0 \\ m_T \sinh \eta\end{pmatrix}= p_T \cosh \eta \, n_J^\mu + \left( m_T-p_T\right)\frac{e^\eta}{2} n_a^\mu + \left( m_T-p_T\right)\frac{e^{-\eta}}{2} n_b^\mu .
\end{equation}
We then change variables from $\xi_a, \xi_b$ to $Q^2,Y$:
\begin{equation}
\xi_a =\frac{e^\eta m_T}{\sqrt{s}} + \frac{e^Y}{\sqrt{s}}\sqrt{p_T^2 + Q^2},
\end{equation}
\begin{equation}
\xi_b = \frac{e^{-\eta} m_T}{\sqrt{s}} + \frac{e^{-Y}}{\sqrt{s}}\sqrt{p_T^2 + Q^2}.
\end{equation}
We can solve the on-shell condition of $p_3$ for $m_T$:
\begin{equation}
m_T = \frac12 \left[ e^\eta \rho_a\Tau_a + e^{-\eta} \rho_b \Tau_b + \sqrt{\left(e^\eta \rho_a \Tau_a +e^{-\eta} \rho_b \Tau_b - 2 p_T \right)^2 + 8 p_T \rho_J \Tau_J \cosh \eta }\right].
\end{equation}
The phase space in the one-jet case is then
\begin{align}
\frac{\d\text{PS}_{ij,1J}}{\d Q^2 \, \d Y\, \d p_T\, \d \eta} =& \frac{\d \hat{\text{PS}}_\text{Born}}{\d Q^2\, \d Y\, \d p_T\, \d \eta} \frac{\Omega_{d-3}\mu_0^{2\varepsilon}}{4 \left(2 \pi \right)^{d-1}}\left(\hat{s}_{ij}\right)^{-1+\varepsilon}\int_0^\pi \d \phi \left(\sin^2 \phi \right)^{-\varepsilon} \nonumber \\
&\int \d \Tau_i\, \Tau_i^{-\varepsilon} \int \d \Tau_j \, \Tau_j^{-\varepsilon} \left( 2 m_T J_{m_T}\right) \frac{x_a x_b}{\xi_a \xi_b}f_i \left( \xi_a\right) f_j \left( \xi_b\right) \delta \left[ \Tau - \hat{\Tau} \left( p_3,p_4\right)\right],
\label{eq:PS1jetgeneral}
\end{align}
where $J_{m_T}$ denotes the Jacobian that arises when removing the
$m_T$ integration.

Finally, we can summarize the structure of the phase space for both the one-jet ($1J$) and the two-jet ($2J$) cases:
\begin{align}
\frac{\d\text{PS}_{ij,nJ}}{\d Q^2 \, \d Y\, \d p_T\, \d \eta} =& \frac{\d \hat{\text{PS}}_\text{Born}}{\d Q^2\, \d Y\, \d p_T\, \d \eta} \frac{\Omega_{d-3}\mu_0^{2\varepsilon}}{4 \left(2 \pi \right)^{d-1}}\left(\hat{s}_{ij}\right)^{-1+\varepsilon}\int_0^\pi \d \phi \left(\sin^2 \phi \right)^{-\varepsilon} \nonumber \\
&\int \d \Tau_i\, \Tau_i^{-\varepsilon} \int \d \Tau_j \, \Tau_j^{-\varepsilon}\, \Phi_{ij,nJ} \left( \Tau_i, \Tau_j, \phi\right)\,  \delta \left[ \Tau - \hat{\Tau} \left( p_3,p_4\right)\right]
\label{eq:PSgeneral}
\end{align}
where the phase space measure $\Phi_{ij,nJ} \left( \Tau_i, \Tau_j,
  \phi\right)$ will be expanded according to the small quantities in
each region.

\subsection{Soft region and non-hemisphere poles}
\label{sec:PSsoft}
In the soft region, all of the components of the emitted gluon
momentum $k$ are soft. Expanding in the soft limit corresponds to
rescaling $\Tau_i \to \lambda \Tau_i$ for all $i$ and taking the limit
$\lambda \to 0$. The expansion of the phase space $\Phi_{ij,nJ}$ as it
appears in Eq.~\eqref{eq:PSgeneral} is
\begin{equation}
\Phi_\text{soft $nJ$} \left(\Tau_a, \Tau_b, \Tau_J \right)= \sum_{n,m,l} \Phi_\text{soft $nJ$}^{(n,m,l)} \, \Tau_a^n\, \Tau_b^m\, \Tau_J^l ,
\label{eq:phiexpsoft}
\end{equation}
where it is understood that in the one-jet case we substitute $\Tau_J\to \Tau'_J$. The expansion coefficients are given in Appendix~\ref{app:PSsoft}.

Due to the fact that the $\Tau_i$ projections are homogeneous in the
soft region, knowing the expansion coefficients is not enough to fully
describe the NLP phase space. We must further study the measurement
function, which in the soft region is expressed by Eq.~\eqref{eq:measurementhemi}. We can split the measurement function into a hemisphere term and a non-hemisphere term, where the latter is made of two pieces: one proportional to $\delta\left(\Tau-\Tau_i \right)$ and the other one proportional to $\delta \left(\Tau-\Tau_m \right)$. We can represent this in the schematic way
\begin{equation}
\delta \left[\Tau-\hat{\Tau}(p_3,p_4) \right] = \mathcal{F}_{ij,\text{hemi}} + \mathcal{F}_{ij,i}+ \mathcal{F}_{ij,m} + \left(i \leftrightarrow j \right),
\end{equation}
where we have defined an hemisphere contribution and two non-hemisphere contributions:
\begin{equation}
\mathcal{F}_{ij,\text{hemi}}  = \Theta \left( \Tau_j-\Tau_i\right) \delta \left( \Tau - \Tau_i\right),
\end{equation}
\begin{equation}
\mathcal{F}_{ij,i} =- \Theta \left( \Tau_j-\Tau_i\right) \Theta \left(\Tau_i - \Tau_m \right) \delta \left( \Tau-\Tau_i\right),
\label{eq:Fiji}
\end{equation}
\begin{equation}
\mathcal{F}_{ij,m} = \Theta \left( \Tau_j-\Tau_i\right) \Theta \left(\Tau_i - \Tau_m \right) \delta \left( \Tau-\Tau_m\right).
\end{equation}
At leading power, the hemisphere terms contain poles, while the non-hemisphere terms are finite. To see why this is the case, we first define the ratios
\begin{equation}
x \equiv \frac{\Tau_j}{\Tau_i} \qquad \qquad z \equiv \frac{\Tau_m}{\Tau_i} = \frac{\hat{s}_{jm}}{\hat{s}_{ij}} + \frac{\hat{s}_{im}}{\hat{s}_{ij}} x -\frac{2 \cos \phi}{\hat{s}_{ij}} \sqrt{\hat{s}_{im} \hat{s}_{jm} x}.
\label{eq:xzdef}
\end{equation}
As we will show in Section~\ref{sec:LP}, the Leading Power (LP) hemisphere cross section is always proportional to
\begin{align}
\sigma^\text{LP}_{\text{soft $ij$,hemi}} &\propto \int_0^\pi \d \phi \left(\sin^2 \phi \right)^{-\varepsilon} \int \d \Tau_i \, \Tau_i^{-1-\varepsilon} \int \d \Tau_j\, \Tau_j^{-1-\varepsilon} \Theta \left( \Tau_j-\Tau_i\right) \delta \left(\Tau-\Tau_i \right)\nonumber \\
&= \frac{\Omega_{d-2}}{\Omega_{d-3}} \Tau^{-1-2\varepsilon} \int \d x \, x^{-1-\varepsilon} \Theta \left(x-1 \right).
\end{align}
The integrand is independent of $\phi$, and the $x$ integral clearly gives a pole when $x\to +\oo$. For the non-hemisphere contributions, the cross sections from the $ij,i$ region and the $ij,m$ region are proportional to
\begin{align}
\sigma^\text{LP}_{\text{soft $ij,i$}} &\propto \int_0^\pi \d \phi \left(\sin^2 \phi \right)^{-\varepsilon} \int \d \Tau_i \, \Tau_i^{-1-\varepsilon} \int \d \Tau_j\, \Tau_j^{-1-\varepsilon} \Theta \left( \Tau_j-\Tau_i\right) \Theta \left( \Tau_i-\Tau_m\right)\delta \left(\Tau-\Tau_i \right)\nonumber \\
&= \Tau^{-1-2\varepsilon}\int_0^\pi \d \phi \left(\sin^2 \phi \right)^{-\varepsilon}  \int \d x\, x^{-1-\varepsilon} \Theta \left( x-1\right) \Theta \left( 1-z\right),
\end{align}
\begin{align}
\sigma^\text{LP}_{\text{soft $ij,m$}} &\propto \int_0^\pi \d \phi \left(\sin^2 \phi \right)^{-\varepsilon} \int \d \Tau_i \, \Tau_i^{-1-\varepsilon} \int \d \Tau_j\, \Tau_j^{-1-\varepsilon} \Theta \left( \Tau_j-\Tau_i\right) \Theta \left( \Tau_i-\Tau_m\right)\delta \left(\Tau- \Tau_m \right)\nonumber \\
&= \Tau^{-1-2\varepsilon}\int_0^\pi \d \phi \left(\sin^2 \phi \right)^{-\varepsilon}  \int \d x\, x^{-1-\varepsilon}\, z^{2\varepsilon} \Theta \left( x-1\right) \Theta \left( 1-z\right).
\end{align}
Both integrals are finite, since the limit $x \to +\oo$ is cut off
from the integral by the constraint $z \le 1$. In principle, there
could be a pole when $z\to0$, but the LP integrand does not contain negative powers of $z$.

This statement is not true anymore at NLP. The soft $ij,i$ contribution will still be finite even at NLP, but the soft $ij,m$ contribution will have a pole. In fact, the power counting is such that a negative power of $z$ does indeed appear at NLP:
\begin{equation}
\sigma^\text{NLP}_\text{soft $ij,m$} \propto \int \d \Tau_i \, \Tau_i^{-2\varepsilon} \delta \left( \Tau - z \Tau_i\right) = \frac{1}{z} \left( \frac{\Tau}{z}\right)^{-2\varepsilon}=\Tau^{-2\varepsilon} z^{-1-2\varepsilon}.
\end{equation}
To better understand this pole, let us investigate in detail the non-hemisphere constraints $\Theta\left( x-1\right) \Theta \left( 1-z\right)$. First, $z\le 1$ can be expressed as
\begin{equation}
z \le 1 \quad \implies \quad \cos \phi \ge \frac{\hat{s}_{jm} + \hat{s}_{im} x - \hat{s}_{ij}}{2\sqrt{\hat{s}_{im} \hat{s}_{jm} x}}\equiv c_\text{min}.
\label{eq:cmin}
\end{equation}
If $c_\text{min} \le -1$, the azimuthal integral is unconstrained. Otherwise, there is a nonzero lower limit in the $\cos \phi$ integral. The two scenarios are respectively represented by the following conditions:
\begin{equation}
c_\text{min} \le -1 \quad \implies \quad 
\begin{cases}
\hat{s}_{ij} \ge \hat{s}_{jm} \\
x \le x_-
\end{cases}
\label{eq:unconstr}
\end{equation}
\begin{equation}
-1\le c_\text{min} \le 1 \quad \implies \quad x_- \le x \le x_+
\label{eq:constr}
\end{equation}
where we have introduced the following limits in the $x$ integral:
\begin{equation}
x_\pm \equiv \frac{\left( \sqrt{\hat{s}_{ij}}\pm\sqrt{\hat{s}_{jm}}\right)^2}{\hat{s}_{im}}.
\label{eq:xpm}
\end{equation}
So far, we have split the $ij,m$ region into a sub-region where the
$\phi$ integral is unconstrained and a region where the $\cos \phi$
integral has a lower limit. Physically, the limit $\cos \phi \to -1$
does not present any singularities, since in that limit $z$ is
strictly positive as can be seen from Eq.~\eqref{eq:xzdef}. Therefore, it makes sense to further split the azimuthal integral into a component that can contain a pole and a finite component:
\begin{equation}
\int_{c_\text{min}}^1 \d \cos \phi = \int_{-1}^1 \d \cos \phi
-\int_{-1}^{c_\text{min}} \d \cos \phi .
\end{equation}
We then define three sub-regions from $ij,m$:
\begin{equation}
\mathcal{F}_{ij,m} = \mathcal{F}_{ij,m,1} +\mathcal{F}_{ij,m,2}+\mathcal{F}_{ij,m,3}
\label{eq:nonhemi123}
\end{equation}
where the constraints in each sub-region are
\begin{equation}
\mathcal{F}_{ij,m,1} =\Theta \left(x-1\right) \Theta \left( \hat{s}_{ij} - \hat{s}_{jm}\right) \Theta \left(x_--x \right) \delta \left( \Tau-\Tau_m\right),
\label{eq:Fijm1}
\end{equation}
\begin{equation}
\mathcal{F}_{ij,m,2} =\Theta \left(x-1\right) \Theta \left( x-x_-\right)\Theta \left( x_+-x\right) \delta \left( \Tau-\Tau_m\right),
\label{eq:Fijm2}
\end{equation}
\begin{equation}
\mathcal{F}_{ij,m,3} =-\Theta \left(x-1\right) \Theta \left( c_\text{min}-\cos \phi\right)\Theta \left( x-x_-\right)\Theta \left( x_+-x\right) \delta \left( \Tau-\Tau_m\right).
\label{eq:Fijm3}
\end{equation}
We have constructed our sub-regions so that the integral $ij,m,3$ is always finite, while $ij,m,1$ and $ij,m,2$ can have a pole. Let us now solve the unconstrained azimuthal integral in the presence of a factor $z^{-1+2\varepsilon}$:
\begin{equation}
 \int_{0}^\pi \d \phi \left( \sin^2 \phi\right)^{-\varepsilon}  z^{-1+2\varepsilon} =\frac{\Omega_{d-2}}{\Omega_{d-3}} \left(\frac{\hat{s}_{ij} }{\hat{s}_{im}}\right)^{1-2\varepsilon} \left| x-x_0\right|^{-1+2\varepsilon},
 \label{eq:azimuthalnonhemi}
\end{equation}
where we have introduced a limit in the $x$ integral where a pole appears:
\begin{equation}
x_0\equiv\frac{\hat{s}_{jm}}{\hat{s}_{im}}.
\label{eq:x0}
\end{equation}
In the special case where $\Tau_m=\Tau'_J$, then the measurement functions produces a factor $\left(z'\right)^{-1+2\varepsilon}$, where
\begin{equation}
z'=\frac{\Tau'_J}{\Tau_i}=z + \mathcal{O} \left( \Tau \right).
\end{equation}
The $\mathcal{O} \left( \Tau \right)$ terms are NNLP, and therefore we do not take them into account.
The relevant integral in $x$, assuming for the time being a generic function $g(x)$ as our integrand, can be expressed in terms of a finite contribution and a pole:
\begin{align}
\int_{x_\text{min}}^{x_\text{max}} \d x \, \frac{g(x)}{ \left|x-x_0 \right|^{1-2\varepsilon}} &=\int_{x_\text{min}}^{x_\text{max}} \d x \frac{g(x) -g(x_0)}{\left|x-x_0\right|}+g(x_0)\int_{x_\text{min}}^{x_\text{max}} \d x \left|x-x_0 \right|^{-1+2\varepsilon}\nonumber \\
&=\int_{x_\text{min}}^{x_\text{max}} \d x \frac{g(x) -g(x_0)}{\left|x-x_0\right|}+\left( \frac{1}{\varepsilon}\right)g\left( x_0\right) \Theta \left(x_\text{max} - x_0 \right)\Theta \left(x_0 - x_\text{min} \right)\nonumber \\
&+g(x_0) \, K_\text{non-hemi} \left(x_0,x_\text{min},x_\text{max} \right).
\label{eq:gintegral}
 \end{align}
The value of the finite integrand depends on the ordering between $x_0$ and the generic integration limits which we named $x_\text{min}$ and $x_\text{max}$:
\begin{align}
K_\text{non-hemi} \left(x_0,x_\text{min}, x_\text{max}\right) &\equiv
\begin{cases}
\log\left[ \left(x_\text{max}-x_0 \right) \left(x_0 - x_\text{min} \right)\right]  \qquad &\text{if} \qquad x_\text{min}\le x_0 \le x_\text{max}, \\
\log \left( \frac{x_0-x_\text{min}}{x_0 - x_\text{max}}\right)\qquad \qquad &\text{if} \qquad x_\text{min}\le x_\text{max} \le x_0, \\
\log \left( \frac{x_\text{max}-x_0}{x_\text{min} - x_0}\right)\qquad \qquad &\text{if} \qquad x_0\le x_\text{min}\le x_\text{max}.
\end{cases}
\end{align}
We notice that the pole is only there if $x_\text{min} \le x_0 \le x_\text{max}$. This condition is satisfied for $\mathcal{F}_{ij,m,1}$ and $\mathcal{F}_{ij,m,2}$ respectively when
\begin{equation}
\mathcal{F}_{ij,m,1}:\quad
\begin{cases}
\hat{s}_{ij} \ge \hat{s}_{jm} \\
1\le x_0 \le x_-
\end{cases}
 \implies \quad
\begin{cases}
\hat{s}_{ij} \ge 4 \hat{s}_{jm} \\
\hat{s}_{jm} \ge \hat{s}_{im}
\end{cases}
\end{equation}
\begin{equation}
\mathcal{F}_{ij,m,2}:\quad
\begin{cases}
x_- \le x_0 \le x_+ \\
x_0\ge 1
\end{cases}\
 \implies \quad
\begin{cases}
\hat{s}_{ij} \le 4 \hat{s}_{jm} \\
\hat{s}_{jm} \ge \hat{s}_{im}
\end{cases}
\end{equation}
Therefore, the non-hemisphere pole term in the cross section is proportional to
\begin{align}
\frac{\d \sigma_\text{soft $ij,m$}^\text{NLP,pole}}{\d Q^2\, \d Y\, \d p_T\, \d \eta} \propto \left(\frac{\Tau^{-2\varepsilon} }{\varepsilon} \right) \left(\frac{\hat{s}_{im}^2}{\hat{s}_{ij}}\right)^\varepsilon\Theta \left(\hat{s}_{jm}-\hat{s}_{im} \right) \left(\hat{s}_{im} \right)^{-1} g(x_0) .
\end{align}
We note that in the case $m=J$, this pole comes from the limit
$\Tau_J^{\prime} \to 0$, and is therefore associated with a soft gluon
emitted close to the hard jet.

This concludes our treatment of the phase space in the soft limit.  To
summarize, we wrote down the expansion of the phase space, then we
analyzed the measurement function and found out that there are new
poles in the non-hemisphere $ij,m$ region corresponding to the limit
$\Tau_m \to 0$. We were able to further split the non-hemisphere
region so as to isolate the poles and separate them from the finite
contributions.  The contribution of these non-hemisphere poles to the
pole cancellation at NLP is an important check of our result.

\subsection{Beam region}
In the beam region, the emitted gluon is collinear to one of the two
initial-state partons. We study explicitly the beam $a$ region, since
the beam $b$ region is related to it by a trivial relabeling $a
\leftrightarrow b$. The quantity that is small in the beam region is
$k_T$, the gluon transverse momentum. In Section~\ref{sec:PS} we
derived a general formula for the phase space, Eq.~\eqref{eq:PSgeneral}. We start from there and make the change of variables
\begin{equation}
\Tau_b = \frac{\sqrt{s} x_a}{\rho_b} \left( \frac{1-z_a}{z_a}\right).
\end{equation}
$z_a$ is the argument of the leading power splitting function, while the transverse momentum of the gluon is
\begin{equation}
k_T = \sqrt{k_a k_b} = \sqrt{\rho_a \rho_b \sqrt{s} x_a \left(\frac{1-z_a}{z_a} \right) } \sqrt{\Tau}.
\end{equation}
An important observation is that in the beam region we expand in
$\sqrt{\Tau}$ rather than in $\Tau$. This might in principle lead to
corrections proportional to $\Tau^{-1/2}$ in the differential cross
section. Such apparent terms cancel upon azimuthal
integration.  Factors of $\sqrt{\Tau}$ are always accompanied by factors of $\cos \phi$, which makes the azimuthal integral vanish.

Upon introducing the momentum fraction $z_a$, the phase space in the beam region is
\begin{align}
\frac{\d\text{PS}_\text{beam $a$}}{\d Q^2\, \d Y\, \d p_T\, \d \eta} =& \frac{\d\hat{\text{PS}}_\text{Born}}{\d Q^2\, \d Y\, \d p_T\, \d \eta} \frac{\Omega_{d-3}\mu_0^{2\varepsilon}}{4(2\pi)^{d-1}} \left(\sqrt{s} x_a \rho_a \right)^{1-\varepsilon}\Tau^{-\varepsilon}\nonumber \\
& \int_0^\pi \d \phi \left( \sin^2 \phi \right)^{-\varepsilon} \int_{x_a}^1 \frac{\d z_a}{z_a^2} \,\left(\frac{1-z_a}{z_a} \right)^{-\varepsilon}\, \Phi_\text{beam $a$} \left( \Tau, z_a, \phi\right).
\label{eq:PSbeama}
\end{align}
The constraint $x_a \le z_a \le 1$ derives from the constraint $0\le \xi_a \le 1$. To be precise, the actual constraint expanded in $\Tau$ is
\begin{equation}
z_a \ge x_a  + \mathcal{O}\left(\sqrt{\Tau} \right).
\label{eq:zalowerlimit}
\end{equation}
Terms of order $\sqrt{\Tau}$ and beyond contribute at NLL, but not at LL.

Finally, as in the soft region we expand the phase space measure:
\begin{equation}
\Phi_\text{beam $a$} \left( \Tau, z_a, \phi\right) = \sum_{n} \Phi_\text{beam $a$}^{(n)}\left(z_a, \phi \right) \Tau^n= \sum_{n,m} \Phi_\text{beam $a$}^{(n,m)}\left( \phi \right) \Tau^n\, \left( 1-z_a\right)^m.
\end{equation}
The relevant expansion coefficients are given in Appendix~\ref{app:PSbeam}.
\subsection{Jet region}
In the jet region, the two gluons in the final state are
collinear. Starting from Eq.~\eqref{eq:PSgeneral}, we choose $ij=Ja$ as Sudakov axes and make the change of variables
\begin{equation}
\Tau_a = \frac{e^{-\eta} p_T z_J}{\rho_a}.
\end{equation}
We note that we could choose $ij=Jb$ as Sudakov axes, and the final result would be the same. The treatment of the jet region follows almost exactly the one of the beam region.
The phase space is
\begin{align}
\frac{\d\text{PS}_\text{jet}}{\d Q^2\, \d Y\, \d p_T\, \d \eta} =& \frac{\d\hat{\text{PS}}_\text{Born}}{\d Q^2\, \d Y\, \d p_T\, \d \eta} \frac{\Omega_{d-3}\mu_0^{2\varepsilon}}{4(2\pi)^{d-1}} \left(2 p_T \rho_J\, \cosh \eta \right)^{1-\varepsilon}\Tau^{-\varepsilon}\nonumber \\
& \int_0^\pi \d \phi \left( \sin^2 \phi \right)^{-\varepsilon} \int_{0}^{1}  \d z_J \, \Theta \left(p_{3T} -p_{4T}\right)\,z_J^{-\varepsilon} \left( 1-z_J \right)^{-\varepsilon}\, \Phi_\text{jet} \left( \Tau, z_J, \phi\right).
\end{align}
The constraint $p_{3T} \ge p_{4T}$ can be explicitly expressed as
\begin{equation}
p_{3T} \ge p_{4T} \qquad \implies \qquad z_J \le \frac12+ \mathcal{O} \left(\sqrt{\Tau} \right)
\label{eq:zJlimit}
\end{equation}
Like for the beam region, terms of order $\sqrt{\Tau}$ and beyond do not contribute to the NLP-LL cross section.

We use the following notation for the expansion of the phase space measure:
\begin{equation}
\Phi_\text{jet} \left( \Tau, z_J, \phi\right) = \sum_{n}
\Phi_\text{jet}^{(n)}\left(z_J, \phi \right) \Tau^n= \sum_{n,m}
\Phi_\text{jet}^{(n,m)}\left( \phi \right) \Tau^n\, z_J^m .
\end{equation}
The expansion coefficients are given in Appendix~\ref{app:PSjet}.

\section{Matrix element expansion}
\label{sec:ME}
For the process of $V+j$ production which we consider in this manuscript, the NLO amplitude
can be taken from~\cite{Ellis:1980wv}. With the full amplitude and having completely
specified all the kinematics in each region, we can proceed to expand the invariants
$s_{ij}$ that appear in the amplitude and hence obtain the expansion of the matrix element
region by region. The notation for the matrix element expansion in the soft region is the
following:
\begin{equation}
\mathcal{M}_\text{soft $nJ$} \left( \Tau_a, \Tau_b, \Tau_J\right)=
\sum_{n,m,l} \mathcal{M}_\text{soft $nJ$}^{(n,m,l)} \Tau_a^n\,
\Tau_b^m\, \Tau_J^l .
\end{equation}
Regarding the beam and jet region, the notation will be
\begin{equation} 
\mathcal{M}_\text{beam $a$} \left( \Tau, z_a, \phi\right) = \sum_n
\mathcal{M}_\text{beam $a$}^{(n)}\left(z_a, \phi \right) \Tau^n =
\sum_{n,m} \mathcal{M}_\text{beam $a$}^{(n,m)} \left(\phi \right)
\Tau^n \left(1-z_a \right)^m ,
\end{equation}
\begin{equation} 
\mathcal{M}_\text{jet} \left( \Tau, z_J, \phi\right) = \sum_n
\mathcal{M}_\text{jet}^{(n)}\left(z_J, \phi \right) \Tau^n =
\sum_{n,m} \mathcal{M}_\text{jet}^{(n,m)} \left(\phi \right) \Tau^n \,
z_J^m .
\end{equation}

The method of expanding the full NLO amplitude is not particularly amenable to a generalization
to more complicated processes where we do not have an analytic
representation of the amplitude.
At leading power, soft and collinear factorization theorems (as
summarized in~\cite{Catani:1999ss} for example)
allow us to predict the first order in the $\Tau$ expansion without knowing the full
NLO amplitude. In fact, a straightforward application of the leading power soft theorem gives us
\begin{equation}
\mathcal{M}_\text{soft $2J$}^{(-1,-1,0)}= \mathcal{M}_\text{soft $1J$}^{(-1,-1,0)}=\left( 4 \pi \alpha_s\right)  \left( C_F -\frac{C_A}{2}\right) 4 \hat{s}_{ab} \mathcal{M}_\text{Born},
\end{equation}
\begin{equation}
\mathcal{M}_\text{soft $2J$}^{(-1,0,-1)}= \mathcal{M}_\text{soft $1J$}^{(-1,0,-1)}=\left( 4 \pi \alpha_s\right) \left(\frac{C_A}{2} \right) 4 \hat{s}_{aJ} \mathcal{M}_\text{Born},
\end{equation}
\begin{equation}
\mathcal{M}_\text{soft $2J$}^{(0,-1,-1)}=\mathcal{M}_\text{soft $1J$}^{(0,-1,-1)}= \left( 4 \pi \alpha_s\right) \left(\frac{C_A}{2} \right) 4 \hat{s}_{bJ} \mathcal{M}_\text{Born}.
\end{equation}
An equally straightforward application of the collinear factorization theorem allows
us to obtain the first terms in the expansion of the matrix element in the beam region
and in the jet region:
\begin{equation}
\mathcal{M}_\text{beam $a$}^{(-1)} =\left( 4 \pi \alpha_s\right) \frac{2 C_F}{\sqrt{s} x_a \rho_a}\left[\frac{1+z_a^2}{1-z_a}-\varepsilon \left(1-z_a \right)\right]  \mathcal{M}_\text{Born},
\label{eq:MELPbeam}
\end{equation}
\begin{equation}
\mathcal{M}_\text{jet}^{(-1)} =\left( 4 \pi \alpha_s\right) \frac{2 C_A}{p_T \rho_J \cosh \eta}\left[\frac{\left( 1-z_J+z_J^2\right)^2}{(1-z_J) z_J} +\mathcal{A}^{(-1)}_\text{jet} (1-z_J)z_J \cos(2 \phi)\right] \mathcal{M}_\text{Born},
\label{eq:MELPjet}
\end{equation}
where the coefficient $\mathcal{A}^{(-1)}_\text{jet}$ does not contribute
to the leading power cross section due to the azimuthal integral
vanishing.  Its explicit form can be obtained using the collinear factorization of the
matrix element presented in~\cite{Catani:1999ss}.

At next-to-leading power, there have been recent efforts towards understanding the
collinear behavior of QCD amplitudes~\cite{Bhattacharya:2018vph}. However,
a factorized formula for the subleading collinear case does not exist yet.
We can only expand the full amplitude in the collinear regions.
Regarding the soft region, a subleading soft theorem in QED
has been known for a long time~\cite{Burnett:1967km}. The extension
to color-ordered QCD amplitudes with the emission of soft gluons does not present any significant issues. The subleading soft theorem reads
\begin{align}
\mathcal{M}^\text{NLO,LP+NLP}_\text{soft} =\left(4 \pi \alpha_s \right)&\sum_{i,j} \Bigg\{-\frac{p_i \cdot p_j}{\left(k \cdot p_i \right)\left( k \cdot p_j\right)}  +\frac{p_i \cdot p_j}{\left(k \cdot p_i \right)\left( k \cdot p_j\right)} k \cdot \frac{\partial}{\partial p_j} -\frac{1}{k \cdot p_i }p_i \cdot \frac{\partial}{\partial p_j}\Bigg\} \nonumber \\
&\Braket{\mathcal{A}_\text{Born}|\vec{T}_i \cdot \vec{T}_j|\mathcal{A}_\text{Born}}.
\label{eq:softtheorem}
\end{align}
Here, $\mathcal{A}_\text{Born}$ indicates the Born amplitude and all the momenta $p_i^\mu$ are incoming. The color factors are in our case
\begin{equation}
\vec{T}_q \cdot \vec{T}_q = \vec{T}_{\bar q} \cdot \vec{T}_{\bar q}=C_F,  \qquad 
\vec{T}_g \cdot \vec{T}_g = C_A, \qquad 
\vec{T}_q \cdot \vec{T}_{\bar q} =\frac{C_A}{2}-C_F, \qquad 
\vec{T}_q \cdot \vec{T}_{g} =\vec{T}_{\bar q} \cdot \vec{T}_{g}= -\frac{C_A}{2}. 
\label{eq:colorfactors}
\end{equation}
The NLO amplitude in the soft limit up to next-to-leading power, expressed as a function of
the NLO invariants $s'_{ij}$ and the Born invariants $s_{ij}$, is
\begin{align}
\mathcal{M}^\text{NLO,LP+NLP}_\text{soft} &=\left(4 \pi \alpha_s \right) \Bigg\{ \left(C_F-\frac{C_A}{2} \right) \frac{4 s'_{12}}{s_{14} s_{24}} +\left(\frac{C_A}{2} \right) \frac{4 s'_{13}}{s_{14} s_{34}} +\left(\frac{C_A}{2} \right)\frac{4 s'_{23}}{s_{24} s_{34}}\Bigg\}\mathcal{M}'_\text{Born}\nonumber \\
&+\left(4 \pi \alpha_s \right)\Bigg\{\frac{1}{s_{14}} \Bigg[ 2 C_F \left(2 s_{12}\frac{\partial}{\partial s_{12}} + s_{13} \left( \frac{\partial}{\partial s_{13}} - \frac{\partial}{\partial s_{23}}\right) \right)\nonumber \\
&+ C_A \left(s_{13} \left(\frac{\partial}{\partial s_{13}} + \frac{\partial}{\partial s_{23}} \right) - s_{12} \left(\frac{\partial}{\partial s_{12}} +\frac{\partial}{\partial s_{23}}\right) \right) \bigg]\nonumber \\
&\frac{1}{s_{24}} \Bigg[ 2 C_F \left(2 s_{12}\frac{\partial}{\partial s_{12}} + s_{23} \left( \frac{\partial}{\partial s_{23}} - \frac{\partial}{\partial s_{13}}\right) \right) \nonumber \\
 &+C_A \left(s_{23} \left(\frac{\partial}{\partial s_{13}} + \frac{\partial}{\partial s_{23}} \right) - s_{12} \left(\frac{\partial}{\partial s_{12}} +\frac{\partial}{\partial s_{13}}\right) \right) \Bigg]\nonumber \\
&+\frac{1}{s_{34}} C_A\left[s_{13} \left(3 \frac{\partial}{\partial s_{13}}  - \frac{\partial}{\partial s_{12}}\right)+ s_{23} \left(3 \frac{\partial}{\partial s_{23}} - \frac{\partial}{\partial s_{12}} \right)\right]\nonumber \\
&+\left( C_F -\frac{C_A}{2}\right) \frac{2 s_{12} s_{34}}{s_{14} s_{24}} \left( \frac{\partial}{\partial s_{13}} +\frac{\partial}{\partial s_{23}}\right)
+\left( \frac{C_A}{2}\right) \frac{2 s_{13} s_{24}}{s_{14} s_{34}} \left( \frac{\partial}{\partial s_{12}}+\frac{\partial}{\partial s_{23}}\right)\nonumber \\
&+\left( \frac{C_A}{2}\right) \frac{2 s_{23} s_{14}}{s_{24} s_{34}} \left( \frac{\partial}{\partial s_{12}}+\frac{\partial}{\partial s_{13}}\right)\Bigg\}\mathcal{M}_\text{Born}.
\label{eq:subleadingsoft}
\end{align}
Eq.~\eqref{eq:subleadingsoft} allows us to extract all of the relevant soft coefficients
and express them in a process independent way in terms of derivatives of the Born matrix
element, without needing to know the full amplitude.  We will see
later that the next-to-soft corrections are sufficient to obtain the
full NLP-LL result, without knowing the exact form of the full amplitude.

We conclude this section by presenting some useful relations between the
beam and jet matrix element expansion coefficients and the soft matrix element
coefficients. The matrix element can be expanded in $\Tau$ and in
$z_a$ or $z_J$. The first orders in the $z_a$ and $z_J$ expansions correspond to
the soft limit of the collinear region, so we are able to map such
coefficients to the soft ones. Since a negative power of $(1-z_a)$ or $z_J$
is the only way that we can produce a pole respectively in the beam region
and in the jet region, having these relations will allow us to check the cancellation
of $\varepsilon^{-1}$ poles. For the beam $a$ region the relations are
\begin{align}
\mathcal{M}_\text{beam $a$}^{(-1,-1)}(\phi) &=\left( \sqrt{s} x_a \rho_a \right)^{-1}\Bigg\{\frac{ \mathcal{M}_\text{soft $2J$}^{(-1,-1,0)}}{\hat{s}_{ab}} +\frac{ \mathcal{M}^{(-1,0,-1)}_\text{soft $2J$}}{\hat{s}_{aJ}}\Bigg\},\\
\vspace{-3mm}\nonumber \\
\mathcal{M}_\text{beam $a$}^{\left(-1/2,-3/2\right)}(\phi)&= \left(\sqrt{s} x_a\rho_a \right)^{-\frac32} \Bigg\{2\sqrt{\frac{\hat{s}_{bJ}}{ \hat{s}_{ab} \hat{s}_{aJ}^3}}  \cos \phi \mathcal{M}_\text{soft $2J$}^{(-1,0,-1)} \Bigg\},\\
\vspace{-3mm} \nonumber \\
\mathcal{M}_\text{beam $a$}^{\left(0,-2\right)}(\phi)&=\left( \sqrt{s} x_a \rho_a\right)^{-2}\Bigg\{\frac{ \mathcal{M}_\text{soft $2J$}^{(0,-1,-1)}}{ \hat{s}_{ab} \hat{s}_{aJ}} 
+\frac{\hat{s}_{bJ}(1+2\cos(2\phi))\mathcal{M}_\text{soft $2J$}^{(-1,0,-1)}}{\hat{s}_{ab} \hat{s}_{aJ}^2 }\Bigg\}, \\
\vspace{-3mm} \nonumber \\
\mathcal{M}_\text{beam $a$}^{\left(0,-1\right)}(\phi)&=\left(\sqrt{s} x_a \rho_a \right)^{-1}\Bigg\{\frac{ \mathcal{M}_\text{soft $2J$}^{(0,-1,0)}}{\hat{s}_{ab}}
+\frac{ \mathcal{M}_\text{soft $2J$}^{(0,0,-1)}}{\hat{s}_{aJ}}-\frac{2 \mathcal{M}_\text{soft $2J$}^{(0,-1,-1)}}{\left( \sqrt{s} x_a \rho_a\right)  \hat{s}_{ab}\hat{s}_{aJ}}+\frac{ \hat{s}_{bJ} \mathcal{M}_\text{soft $2J$}^{(-1,-1,1)}}{\hat{s}_{ab}^2}
\nonumber \\
& -\frac{2 \hat{s}_{bJ}(1+2 \cos(2 \phi)) \mathcal{M}^{(-1,0,-1)}_\text{soft $2J$}}{\left( \sqrt{s} x_a \rho_a\right) \hat{s}_{ab}\hat{s}_{aJ}^2}
+\frac{ \hat{s}_{bJ}(1+2 \cos(2 \phi)) \mathcal{M}^{(-1,1,-1)}_\text{soft $2J$}}{\hat{s}_{aJ}^2}\Bigg\}.
\end{align}
For the beam $b$ region, the relations are the same upon relabeling $a \leftrightarrow b$, $Y \leftrightarrow -Y$, $\eta \leftrightarrow - \eta$. The relations between the matrix element in the jet region and the soft region are
\begin{align}
\mathcal{M}_\text{jet}^{(-1,-1)}(\phi) &=\left( 2 p_T \rho_J \cosh \eta\right)^{-1}\Bigg\{\frac{ \mathcal{M}_\text{soft $1J$}^{(-1,0,-1)}}{\hat{s}_{aJ}} +\frac{ \mathcal{M}_\text{soft $1J$}^{(0,-1,-1)}}{\hat{s}_{bJ}}\Bigg\},\\
\vspace{-3mm} \nonumber \\
\mathcal{M}_\text{jet}^{\left(-1/2,-3/2\right)}(\phi)&=\left(2 p_T \rho_J \cosh \eta \right)^{-\frac32} \Bigg\{2\sqrt{\frac{\hat{s}_{ab} }{ \hat{s}_{aJ}\hat{s}_{bJ}^3}}\cos \phi  \mathcal{M}_\text{soft $1J$}^{(0,-1,-1)}\Bigg\}, \\
\vspace{-3mm} \nonumber \\
\mathcal{M}_\text{jet}^{\left(0,-2\right)}(\phi)&=\left( 2 p_T \rho_J \cosh \eta \right)^{-2}\Bigg\{\frac{\mathcal{M}_\text{soft $1J$}^{(-1,-1,0)}}{\hat{s}_{aJ} \hat{s}_{bJ}}+\frac{\hat{s}_{ab} (1+2 \cos (2\phi))  \mathcal{M}_\text{soft $1J$}^{(0,-1,-1)}}{ \hat{s}_{aJ}\hat{s}_{bJ}^2}\Bigg\},\\
\vspace{-3mm} \nonumber \\
\mathcal{M}_\text{jet}^{\left(0,-1\right)}(\phi)&= \left(2 p_T \rho_J \cosh \eta \right)^{-1}\Bigg\{\frac{  \mathcal{M}_\text{soft $1J$}^{(-1,0,0)}}{\hat{s}_{aJ}}+\frac{ \mathcal{M}_\text{soft $1J$}^{(0,-1,0)}}{\hat{s}_{bJ}}-
 \frac{\hat{s}_{ab}(3 \cos (2 \phi )+1)\mathcal{M}_\text{soft $1J$}^{(0,-1,-1)}  }{\left( 2 p_T \rho_J \cosh \eta\right) 2 \hat{s}_{aJ} \hat{s}_{bJ}^2}
 \nonumber \\
 &+\frac{ \hat{s}_{ab} (2 \cos (2 \phi )+1)\mathcal{M}_\text{soft $1J$}^{(1,-1,-1)}}{ \hat{s}_{bJ}^2}
 +\frac{ \hat{s}_{ab}\mathcal{M}_\text{soft $1J$}^{(-1,1,-1)}}{\hat{s}_{aJ}^2}\Bigg\}.
\end{align}
\section{Leading power cross section}
\label{sec:LP}
In this section we reproduce the leading-power cross section in the
small-$\Tau$ limit as a check on our approach.  To obtain this result,
we multiply the leading-power matrix element by the leading-power
phase space.  We arrange the calculation into beam, jet and soft
functions to match results in the literature.  For simplicity we omit
an explicit discussion of the hard function, which matches exactly the
virtual corrections to the cross section in dimensional regularization.

\subsection{Soft function}
\label{sec:softLP}
To compute the soft function contribution we first consider the
integrand in the soft region:
\begin{align}
\frac{\d\sigma_\text{soft}^\text{LP}}{\d Q^2\, \d Y\, \d p_T\, \d \eta} =&\frac{\d \hat{\text{PS}}_\text{Born}}{\d Q^2\, \d Y\, \d p_T\, \d \eta} \left(\frac{\alpha_s}{4\pi} \right)\frac{\left(4 \pi \mu_0^2\right)^\varepsilon}{ \sqrt{\pi} \Gamma \left(\frac12-\varepsilon \right)}\left(\hat{s}_{ij}\right)^{-1+\varepsilon}\int_0^\pi \d \phi \left(\sin^2 \phi \right)^{-\varepsilon} \nonumber \\
&\int \d \Tau_i\, \Tau_i^{-\varepsilon} \int \d \Tau_j \, \Tau_j^{-\varepsilon}\,  f_q \left(x_a \right) f_{\bar{q}} \left(x_b\right)\delta \left[ \Tau - \hat{\Tau} \left( p_3,p_4\right)\right]\nonumber \\
&\Bigg\{ \frac{\mathcal{M}_\text{soft}^{(-1,-1,0)}}{\Tau_a \Tau_b}+\frac{\mathcal{M}_\text{soft}^{(-1,0,-1)}}{\Tau_a \Tau_J}+\frac{\mathcal{M}_\text{soft}^{(0,-1,-1)}}{\Tau_b \Tau_J}\Bigg\}.
\end{align}
The superscripts on the matrix element structures indicate the powers of
$\Tau_i$ that appear in the denominator for that term. At leading power, there is no difference between the two-jet
parametrization and the one-jet parametrization for both the phase space and matrix element.
The structure of the measurement function of
Eq.~\eqref{eq:measurementhemi} in the soft region requires us to
choose the two Sudakov axes that appear in the hemisphere
decomposition.  We make the following choices:
\begin{itemize}
\item $ij=ab$ will be used for $\mathcal{M}_\text{soft}^{(-1,-1,0)}$;
\item $ij=aJ$ will be used for $\mathcal{M}_\text{soft}^{(-1,0,-1)}$;
\item $ij=bJ$ will be used for $\mathcal{M}_\text{soft}^{(0,-1,-1)}$.
\end{itemize}
Due to the symmetric configuration of the integrand, we can express the result in terms of the $ij$ hemisphere contribution and the $ij,i$ and $ij,m$ non-hemisphere contributions, following the decomposition of the measurement function from Section~\ref{sec:PSsoft}. The contributions to the differential cross section are
\begin{align}
\frac{\d\sigma_\text{soft $ij$ hemi}^\text{LP}}{\d Q^2\, \d Y\, \d p_T\, \d \eta} &=\frac{\d  \sigma_\text{Born}}{\d Q^2\, \d Y\, \d p_T\, \d \eta} \left(4\vec{T}_i \cdot \vec{T}_{j}\right)\left(\frac{\alpha_s}{4\pi} \right)\frac{\left(e^{\gamma_E} \mu^2\right)^\varepsilon}{ \sqrt{\pi} \Gamma \left(\frac12-\varepsilon \right)}\left(\hat{s}_{ij}\right)^{\varepsilon}\int_0^\pi \d \phi \left(\sin^2 \phi \right)^{-\varepsilon} \nonumber \\
&\int \d \Tau_i\, \Tau_i^{-1-\varepsilon} \int \d \Tau_j \, \Tau_j^{-1-\varepsilon}\Theta \left(\Tau_j-\Tau_i \right) \delta \left( \Tau-\Tau_i\right)\nonumber \\
&=\frac{\d  \sigma_\text{Born}}{\d Q^2\, \d Y\, \d p_T\, \d \eta} \left(\vec{T}_i \cdot \vec{T}_j\right)\left(\frac{\alpha_s}{4\pi} \right) \left[\frac{8}{\sqrt{\hat{s}_{ij}}\mu} \mathcal{L}_1 \left( \frac{\Tau}{\sqrt{\hat{s}_{ij}}\mu}\right) -\frac{\pi^2}{6}\delta \left( \Tau\right)\right],\\ \vspace{2mm} \nonumber \\
\frac{\d\sigma_\text{soft $ij,i$}^\text{LP}}{\d Q^2\, \d Y\, \d p_T\, \d \eta} =&\frac{\d \sigma_\text{Born}}{\d Q^2\, \d Y\, \d p_T\, \d \eta} \left(4\vec{T}_i \cdot \vec{T}_j\right)\left(\frac{\alpha_s}{4\pi} \right)\frac{\left(e^{\gamma_E} \mu^2\right)^\varepsilon}{ \sqrt{\pi} \Gamma \left(\frac12-\varepsilon \right)}\left(\hat{s}_{ij}\right)^{\varepsilon} \Tau^{-1-2\varepsilon} \nonumber \\
&\int_0^\pi \d \phi \left(\sin^2 \phi \right)^{-\varepsilon} \int \d x \, x^{-1-\varepsilon}\Theta \left(x-1\right) \Theta \left(1-z \right) ,
\label{eq:LPiji}\\\vspace{2mm}\nonumber \\
\frac{\d\sigma_\text{soft $ij,m$}^\text{LP}}{\d Q^2\, \d Y\, \d p_T\, \d \eta} =&-\frac{\d \sigma_\text{Born}}{\d Q^2\, \d Y\, \d p_T\, \d \eta} \left(4\vec{T}_i \cdot \vec{T}_j\right)\left(\frac{\alpha_s}{4\pi} \right)\frac{\left(e^{\gamma_E} \mu^2\right)^\varepsilon}{ \sqrt{\pi} \Gamma \left(\frac12-\varepsilon \right)}\left(\hat{s}_{ij}\right)^{\varepsilon} \Tau^{-1-2\varepsilon}  \nonumber \\
&\int_0^\pi \d \phi \left(\sin^2 \phi \right)^{-\varepsilon} \int \d x \, x^{-1-\varepsilon}\, z^{2\varepsilon}\, \Theta \left(x-1 \right) \Theta \left(1-z \right) .
\end{align}
The color factors $\vec{T}_i \cdot \vec{T}_j$ have been introduced in
Eq.~\eqref{eq:colorfactors}.  $i$ and $j$ can indicate either a quark, an anti-quark or a gluon.
We have used the following $\varepsilon$ expansions:
\begin{equation}
\frac{e^{\varepsilon \gamma_E} }{\Gamma(1-\varepsilon)} = 1-\frac{\pi^2 \varepsilon^2}{12} + \mathcal{O} \left(\varepsilon^3 \right),
\label{eq:LPijm}
\end{equation}
\begin{align}
 \left(\frac{1}{\sqrt{\hat{s}_{ij}}\mu} \right) \left( \frac{ \Tau}{\sqrt{\hat{s}_{ij}}\mu}\right)^{-1-2\varepsilon} \to - \frac{\delta \left( \Tau\right)}{2\varepsilon} +  \frac{1}{\sqrt{\hat{s}_{ij}}\mu}\mathcal{L}_0 \left( \frac{\Tau}{\sqrt{\hat{s}_{ij}}\mu} \right)- \frac{2\varepsilon}{\sqrt{\hat{s}_{ij}}\mu}  \mathcal{L}_1 \left( \frac{\Tau}{\sqrt{\hat{s}_{ij}}\mu} \right),
\end{align}
where we have defined the standard plus distributions:
\begin{equation}
\mathcal{L}_n(x) = \left[\frac{\log^{n}(x)}{x} \right]_+ .
\end{equation}  
The sum of the non-hemisphere
integrals of Eqs.~(\ref{eq:LPiji}--\ref{eq:LPijm}) can be expanded in
$\varepsilon$ and cast in the same form as in~\cite{Jouttenus:2011wh}.
We therefore reproduce the leading-power soft function known in the literature.
\subsection{Beam function}
\label{sec:beamLP}
 It is straightforward to obtain the differential cross section using the leading-power phase space and matrix element in the beam region:
\begin{align}
\frac{\d \sigma^\text{LP}_\text{beam $a$}}{\d Q^2\, \d Y\, \d p_T\,  \d \eta\, \d \Tau} =& \frac{\d\hat{\sigma}_\text{Born}}{\d Q^2\, \d Y \, \d p_T \, \d \eta} \left( \frac{C_F \alpha_s}{2\pi}\right)  \frac{e^{\varepsilon \gamma_E} }{\Gamma(1-\varepsilon)} \left(\frac{\sqrt{s} x_a \rho_a}{\mu^2} \right) \left( \frac{\sqrt{s} x_a \rho_a \Tau}{\mu^2}\right)^{-1-\varepsilon}\nonumber \\
&   \int_{x_a}^1 \frac{\d z_a}{z_a}\, \left(\frac{1-z_a}{z_a} \right)^{-\varepsilon}  f_i\left(\frac{x_a}{z_a}\right) f_j\left(x_b\right)  \left[ \frac{1+z_a^2}{1-z_a} -\varepsilon(1-z_a)\right],
\end{align}
where $\hat{\sigma}_\text{Born}$ indicates the Born-level partonic
cross section with PDFs removed. We use the following $\varepsilon$ expansions:
\begin{align}
 \left(\frac{\sqrt{s} x_a \rho_a}{\mu^2} \right) \left( \frac{\sqrt{s} x_a \rho_a \Tau}{\mu^2}\right)^{-1-\varepsilon} \to &- \frac{\delta \left( \Tau\right)}{\varepsilon} + \left(\frac{\sqrt{s} x_a \rho_a}{\mu^2} \right) \mathcal{L}_0 \left( \frac{\sqrt{s} x_a \rho_a \Tau}{\mu^2} \right)\nonumber \\
 &- \varepsilon \left(\frac{\sqrt{s} x_a \rho_a}{\mu^2} \right) \mathcal{L}_1 \left( \frac{\sqrt{s} x_a \rho_a \Tau}{\mu^2} \right),
\end{align}
\begin{equation}
\left( 1-z_a\right)^{-1-\varepsilon} z_a^\varepsilon \to -\frac{\delta(1-z_a)}{\varepsilon}+\mathcal{L}_0(1-z_a) - \varepsilon \left[ \mathcal{L}_1(1-z_a)- \log z_a\mathcal{L}_0(1-z_a)\right],
\end{equation}
The finite part of the cross section takes the form
\begin{align}
\frac{\d \sigma^\text{LP}_{\text{beam $a$}}}{\d Q^2\, \d Y\, \d p_T\,  \d \eta\, \d \Tau} =&\frac{\d\hat{\sigma}_\text{Born}}{\d Q^2\, \d Y \, \d p_T \, \d \eta} \left( \frac{C_F \alpha_s}{2\pi}\right) \int_{x_a}^1 \frac{\d z_a}{z_a}f_i\left(\frac{x_a}{z_a}\right) f_j\left(x_b\right)\nonumber \\
& \Bigg\{\delta\left(\Tau \right)\left[-\delta\left(1-z_a \right) \frac{\pi^2}{6} + \mathcal{L}_1\left(1-z_a \right)(1+z_a^2) + \left(1-z_a-\frac{1+z_a^2}{1-z_a} \log z_a\right)\right]\nonumber \\
&+ \left(\frac{\sqrt{s} x_a \rho_a}{\mu^2} \right) \mathcal{L}_0 \left( \frac{\sqrt{s} x_a \rho_a \Tau}{\mu^2} \right)\mathcal{L}_0\left(1-z_a \right)(1+z_a^2) \nonumber \\
&+ 2 \delta(1-z_a) \left(\frac{\sqrt{s} x_a \rho_a}{\mu^2} \right) \mathcal{L}_1 \left( \frac{\sqrt{s} x_a \rho_a \Tau}{\mu^2} \right) \Bigg\}.
\end{align}
This corresponds exactly to the quark beam function contribution to the
cross section in the literature~\cite{Stewart:2010qs}.

\subsection{Jet function}
\label{sec:jetLP}
Combining the leading-power phase space and matrix element in the jet region we obtain the differential cross section:
\begin{align}
\frac{\d \sigma^\text{LP}_\text{jet}}{\d Q^2 \, \d Y\, \d p_T\, \d \eta \d \Tau} &=\frac{\d \sigma_\text{Born}}{\d Q^2 \, \d Y\, \d p_T\, \d \eta}\frac{e^{ \varepsilon \gamma_E}  }{ \Gamma \left(1-\varepsilon \right)}\left(\frac{ C_A \alpha_s}{ \pi}\right) \left(\frac{2 p_T \rho_J \cosh \eta}{\mu^2} \right)    \nonumber \\
&\left( \frac{2 p_T \rho_J \cosh \eta\, \Tau}{\mu^2} \right)^{-1-\varepsilon} \int_0^{\frac12} \d z_J  z_J^{-\varepsilon}(1-z_J)^{-\varepsilon}  \frac{\left( 1-z_J + z_J^2\right)^2}{z_J (1-z_J)}.
\end{align}
We use the following $\varepsilon$ expansion:
\begin{align}
 \left(\frac{2 p_T \rho_J \cosh \eta}{\mu^2} \right) \left( \frac{2 p_T \rho_J \cosh \eta \Tau}{\mu^2}\right)^{-1-\varepsilon} \to &- \frac{\delta \left( \Tau\right)}{\varepsilon} + \left(\frac{2 p_T \rho_J \cosh \eta}{\mu^2} \right) \mathcal{L}_0 \left( \frac{2 p_T \rho_J \cosh \eta \Tau}{\mu^2} \right)\nonumber \\
 &- \varepsilon \left(\frac{2 p_T \rho_J \cosh \eta}{\mu^2} \right) \mathcal{L}_1 \left( \frac{2 p_T \rho_J \cosh \eta \Tau}{\mu^2} \right).
\end{align}
We can also perform the integral in $z_J$:
\begin{equation}
\int_0^{\frac12}\d z_J  z_J^{-1-\varepsilon}(1-z_J)^{-1-\varepsilon} \left(1-z_J+z_J^2\right)^2 = -\frac{1}{\varepsilon} - \frac{11}{12} + \varepsilon \left(-\frac{67}{36} + \frac{\pi^2}{6} \right).
\end{equation}
The finite part of the differential cross section becomes
\begin{align}
\frac{\d \sigma^\text{LP}_\text{jet}}{\d Q^2 \, \d Y\, \d p_T\, \d \eta \d \Tau} &=\frac{\d \sigma_\text{Born}}{\d Q^2 \, \d Y\, \d p_T\, \d \eta}\left(\frac{  \alpha_s C_A}{\pi} \right) \Bigg\{ \delta(\Tau) \left[\frac{67}{36} -\frac{\pi^2}{4}\right] \nonumber \\
&- \frac{11}{12}\left(\frac{2 p_T \rho_J \cosh \eta}{\mu^2} \right) \mathcal{L}_0 \left( \frac{2 p_T \rho_J \cosh \eta \Tau}{\mu^2} \right) \nonumber \\
&+  \left(\frac{2 p_T \rho_J \cosh \eta}{\mu^2} \right) \mathcal{L}_1 \left( \frac{2 p_T \rho_J \cosh \eta \Tau}{\mu^2} \right)\Bigg\}.
\end{align}
This corresponds to the gluon contribution to the jet function with
$n_f$ set to zero~\cite{Becher:2009th}.

\section{Next-to-leading power cross section}
\label{sec:NLP}

In this section we derive the cross section at next-to-leading
power.  We organize the calculation using the previously-defined beam,
jet and soft regions.  For each region we discuss the terms that enter the NLP cross section,
focusing on the leading logarithmic contributions first, characterized by the presence of a pole, and
discussing the finite next-to-leading logarithmic contributions later. Eventually, the final result will take the form
\begin{align}
\frac{\d \sigma^\text{NLP}}{\d Q^2\, \d Y\, \d p_T\, \d \eta\, \d \Tau} &= \frac{\d \hat{\text{PS}}_\text{Born}}{\d Q^2\, \d Y\, \d p_T\, \d \eta} \left( \frac{\alpha_s}{4\pi}\right) \sum_\alpha \bigg\{C^\text{LL}_\alpha \log \frac{\Tau}{Q_\alpha} + C^\text{NLL}_\alpha \bigg\},
\label{eq:NLPXS}
\end{align}
where the index $\alpha$ runs over all the regions: beam $a$, beam $b$, jet, soft $ij$ hemi, soft $ij$ non-hemi.
\subsection{Soft region}
\label{sec:NLPsoft}
We start our treatment by defining the product of the soft matrix element times the soft phase space:
\begin{equation}
\mathcal{S}_{nJ} \left( \Tau_a, \Tau_b, \Tau_J\right) \equiv \left(\frac{1}{4\pi \alpha_s} \right)\mathcal{M}_\text{soft $nJ$} \left( \Tau_a, \Tau_b, \Tau_J\right)\Phi_\text{soft $nJ$} \left( \Tau_a, \Tau_b, \Tau_J\right).
\end{equation}
The NLP cross section in the soft region is, using $i$ and $j$ as reference axes,
\begin{align}
\frac{\d\sigma_\text{soft}^\text{NLP}}{\d Q^2\, \d Y\, \d p_T\, \d \eta} =&\frac{\d \hat{\text{PS}}_\text{Born}}{\d Q^2\, \d Y\, \d p_T\, \d \eta} \left(\frac{\alpha_s}{4\pi} \right)\frac{\left(4 \pi \mu_0^2\right)^\varepsilon}{ \sqrt{\pi} \Gamma \left(\frac12-\varepsilon \right)}\left(\hat{s}_{ij}\right)^{-1+\varepsilon}\int_0^\pi \d \phi \left(\sin^2 \phi \right)^{-\varepsilon} \nonumber \\
&\int \d \Tau_i\, \Tau_i^{-\varepsilon} \int \d \Tau_j \, \Tau_j^{-\varepsilon}\,  \delta \left[ \Tau - \hat{\Tau} \left( p_3,p_4\right)\right]\Bigg\{ \frac{\mathcal{S}_{nJ}^{(-1,0,0)}}{\Tau_a}+\frac{\mathcal{S}_{nJ}^{(0,-1,0)}}{\Tau_b}+\frac{\mathcal{S}_{nJ}^{(0,0,-1)}}{\Tau_J}\nonumber \\
&+ \frac{\mathcal{S}_{nJ}^{(-1,-1,1)}\Tau_J}{\Tau_a \Tau_b}+ \frac{\mathcal{S}_{nJ}^{(-1,1,-1)}\Tau_b}{\Tau_a \Tau_J}+ \frac{\mathcal{S}_{nJ}^{(1,-1,-1)} \Tau_a}{\Tau_b \Tau_J}\Bigg\},
\label{eq:Sfunctions}
\end{align}
where the measurement function determines whether the two-jet
parametrization or the one-jet parametrization should be used.  The
superscripts on the $\mathcal{S}_{nJ}$ denote the powers in the
$\Tau_i$ expansions; for example, $\mathcal{S}_{nJ}^{(-1,-1,1)}$
denotes the coefficient of the term with $\Tau_a$ and $\Tau_b$ in the
denominator, and $\Tau_J$ in the numerator.  The $n$ in the subscript
denotes whether the 1-jet or 2-jet parameterization is used. The
measurement function in the hemisphere decomposition requires us to
choose a pair of Sudakov axes $ij$ for each term in the integrand,
just as at leading power. This choice will not affect the final
result, and we choose the most symmetric configuration:
\begin{itemize}
\item $ij=ab$ will be used for $\mathcal{S}_{nJ}^{(-1,-1,1)}$, half of $\mathcal{S}_{nJ}^{(-1,0,0)}$ and half of $\mathcal{S}_{nJ}^{(0,-1,0)}$;
\item $ij=aJ$ will be used for $\mathcal{S}_{nJ}^{(-1,1,-1)}$, half of $\mathcal{S}_{nJ}^{(-1,0,0)}$ and half of $\mathcal{S}_{nJ}^{(0,0,-1)}$;
\item $ij=bJ$ will be used for $\mathcal{S}_{nJ}^{(1,-1,-1)}$, half of $\mathcal{S}_{nJ}^{(0,-1,0)}$ and half of $\mathcal{S}_{nJ}^{(0,0,-1)}$.
\end{itemize}
In order to compute the hemisphere cross section, we need the following integrals:
\begin{equation}
\int_0^\pi \d \phi \left( \sin^2 \phi\right)^{-\varepsilon} \int \d \Tau_i \,  \int \d \Tau_j \, \Tau_i^{-1-\varepsilon}\, \Tau_j^{-\varepsilon}\Theta \left( \Tau_j-\Tau_i\right) \delta \left( \Tau-\Tau_i\right) =\frac{\Omega_{d-2}}{\Omega_{d-3}}\Tau^{-2\varepsilon} \left[-1+\mathcal{O}\left( \varepsilon\right)\right],
\end{equation}
\begin{equation}
\int_0^\pi \d \phi \left( \sin^2 \phi\right)^{-\varepsilon} \int \d \Tau_i \,  \int \d \Tau_j \, \Tau_i^{-\varepsilon}\, \Tau_j^{-1-\varepsilon}\Theta \left( \Tau_j-\Tau_i\right) \delta \left( \Tau-\Tau_i\right) =\frac{\Omega_{d-2}}{\Omega_{d-3}}\Tau^{-2\varepsilon} \left(\frac{1}{\varepsilon}\right),
\end{equation}
\begin{align}
&\int_0^\pi \d \phi \left( \sin^2 \phi\right)^{-\varepsilon} \int \d \Tau_i \,  \int \d \Tau_j \, \Tau_i^{-1-\varepsilon}\, \Tau_j^{-1-\varepsilon} \left(\frac{\hat{s}_{jm}}{\hat{s}_{ij}} \Tau_i + \frac{\hat{s}_{im}}{\hat{s}_{ij}}\Tau_j -2 \cos \phi\frac{\sqrt{\hat{s}_{im}\hat{s}_{jm} \Tau_i \Tau_j}}{\hat{s}_{ij}} \right) \nonumber \\
&\Theta \left( \Tau_j-\Tau_i\right) \delta \left( \Tau-\Tau_i\right)=\frac{\Omega_{d-2}}{\Omega_{d-3}}\Tau^{-2\varepsilon} \left[\frac{1}{\varepsilon} \frac{\hat{s}_{jm}}{\hat{s}_{ij}}-\frac{\hat{s}_{im}}{\hat{s}_{ij}}+\mathcal{O}\left( \varepsilon\right)\right].
\end{align}
The results shown are straightforward to derive by direct
integration.  We now have all the ingredients needed to compute the
hemisphere cross section. We sum over all the hemispheres and obtain
\begin{align}
\frac{\d\sigma^\text{NLP,LL}_\text{soft hemi}}{\d Q^2\, \d Y\, \d p_T\, \d \eta} =&\frac{\d \hat{\text{PS}}_\text{Born}}{\d Q^2\, \d Y\, \d p_T\, \d \eta} \left( \frac{\alpha_s}{4 \pi}\right) \frac{\left(4 \pi \mu_0^2\right)^{\varepsilon}}{ \Gamma \left(1-\varepsilon \right)}\Tau^{-2\varepsilon}\left(\frac{1}{\varepsilon}\right)\Bigg\{ 
 \frac{\mathcal{S}^{(-1,0,0)}_{2J} +\mathcal{S}^{(0,-1,0)}_{2J}}{2 \left( \hat{s}_{ab}\right)^{1-\varepsilon}}\nonumber \\
&+ \frac{\mathcal{S}^{(-1,0,0)}_{1J} +\mathcal{S}^{(0,0,-1)}_{2J}}{2 \left( \hat{s}_{aJ}\right)^{1-\varepsilon}} 
+ \frac{\mathcal{S}^{(0,-1,0)}_{1J} +\mathcal{S}^{(0,0,-1)}_{2J}}{2 \left( \hat{s}_{bJ}\right)^{1-\varepsilon}} 
+\frac{\mathcal{S}_{2J}^{(-1,-1,1)} \hat{s}_{bJ}+\mathcal{S}_{2J}^{(-1,-1,1)} \hat{s}_{aJ}}{\left( \hat{s}_{ab}\right)^{2-\varepsilon}}\nonumber \\
& +\frac{\mathcal{S}_{2J}^{(-1,1,-1)} \hat{s}_{bJ}+\mathcal{S}_{1J}^{(-1,1,-1)} \hat{s}_{ab}}{\left( \hat{s}_{aJ}\right)^{2-\varepsilon}} 
+\frac{\mathcal{S}_{2J}^{(1,-1,-1)} \hat{s}_{aJ}+\mathcal{S}_{1J}^{(1,-1,-1)} \hat{s}_{ab}}{\left( \hat{s}_{bJ}\right)^{2-\varepsilon}}\Bigg\} .
\end{align}

To compute the non-hemisphere NLP soft function, we note that in Section~\ref{sec:PSsoft} we analyzed the structure of the constraints and described a way to separate the LL structures from the NLL ones. Summing over all contributions, the NLP-LL cross section is
\begin{align}
\frac{\d\sigma^\text{NLP,LL}_\text{soft non-hemi}}{\d Q^2\, \d Y\, \d p_T\, \d \eta} =&\frac{\d \hat{\text{PS}}_\text{Born}}{\d Q^2\, \d Y\, \d p_T\, \d \eta} \left( \frac{\alpha_s}{4 \pi}\right) \frac{\left(4 \pi \mu_0^2\right)^{\varepsilon}}{ \Gamma \left(1-\varepsilon \right)}\Tau^{-2\varepsilon}\left(\frac{1}{\varepsilon}\right)\nonumber \\
\Bigg\{&\left( \frac{\mathcal{S}_{1J}^{(-1,0,0)}}{2\hat{s}_{aJ}}+\frac{\mathcal{S}_{1J}^{(0,-1,0)}}{2\hat{s}_{bJ}}\right) \left[ \left( \frac{\hat{s}_{aJ}^2}{\hat{s}_{ab}}\right)^\varepsilon \Theta \left( \hat{s}_{bJ} - \hat{s}_{aJ}\right) +\left(\frac{\hat{s}_{bJ}^2}{\hat{s}_{ab}}\right)^{\varepsilon} \Theta \left( \hat{s}_{aJ} -\hat{s}_{bJ}\right)\right] \nonumber \\&\left( \frac{\mathcal{S}_{2J}^{(-1,0,0)}}{2\hat{s}_{ab}}+\frac{\mathcal{S}_{2J}^{(0,0,-1)}}{2\hat{s}_{bJ}}\right) \left[ \left( \frac{\hat{s}_{ab}^2}{\hat{s}_{aJ}}\right)^\varepsilon \Theta \left( \hat{s}_{bJ} - \hat{s}_{ab}\right) +\left(\frac{\hat{s}_{bJ}^2}{\hat{s}_{aJ}}\right)^{\varepsilon} \Theta \left( \hat{s}_{ab} -\hat{s}_{bJ}\right)\right] \nonumber \\
&\left( \frac{\mathcal{S}_{2J}^{(0,-1,0)}}{2\hat{s}_{ab}}+\frac{\mathcal{S}_{2J}^{(0,0,-1)}}{2\hat{s}_{aJ}}\right) \left[ \left( \frac{\hat{s}_{ab}^2}{\hat{s}_{bJ}}\right)^\varepsilon \Theta \left( \hat{s}_{aJ} - \hat{s}_{ab}\right) +\left(\frac{\hat{s}_{aJ}^2}{\hat{s}_{bJ}}\right)^{\varepsilon} \Theta \left( \hat{s}_{ab} -\hat{s}_{aJ}\right)\right] \Bigg\} .
\end{align}
%
\subsection{Beam region}
\label{sec:NLPbeam}
Like in the soft region, we define the product of matrix element and
phase space. This time, we integrate inclusively over the azimuthal
angle, since no observable constrains this variable:
\begin{equation}
\mathcal{B}_a \left( \Tau,z_a\right) \equiv \left(\frac{1}{4\pi \alpha_s} \right)\frac{\Omega_{d-3}}{\Omega_{d-2}}\int_0^\pi \d \phi \left( \sin^2\phi\right)^{-\varepsilon} \mathcal{M}_\text{beam $a$} \left(\Tau,z_a,\phi \right) \Phi_\text{beam $a$} \left( \Tau, z_a, \phi\right).
\end{equation}
We note that the $z_a$ expansion of $\mathcal{B}_a^{(0)}(z_a)$ is
\begin{equation}
\mathcal{B}_a^{(0)}(z_a)= \frac{\mathcal{B}_a^{(0,-2)} +\mathcal{B}_a^{(0,-1)}(1-z_a)+\mathcal{B}_a^{(0,0)}(1-z_a)^2+\dots}{(1-z_a)^2} .
\label{eq:Bexpansion}
\end{equation}
This means that the first two terms in the $z_a$ expansion of
$\mathcal{B}_a^{(0)}(z_a)$ will contribute to the LL cross section,
while the remaining terms are finite. In particular, we notice that
the first term in Eq.~(\eqref{eq:Bexpansion}) has the form of a
power-law divergence. This peculiar behavior was also observed in Refs.~\cite{Bhattacharya:2018vph} and~\cite{Beneke:2019kgv} at next-to-leading power in the case of multiple collinear directions. The finite contribution coming from this term, however, is a NLP-NLL contribution, as it is not associated with a pole (and therefore a leading log).
In order to extract the pole, we sum and subtract the first two terms in the expansion:
\begin{align}
\int_{x_a}^1  \frac{\d z_a}{z_a^{2-\varepsilon}}\,  \mathcal{B}_a^{(0)}(z_a) &=\int_{x_a}^1  \frac{\d z_a}{z_a^2} \left[\mathcal{B}_a^{(0)}(z_a) -\frac{\mathcal{B}_a^{(0,-2)}}{(1-z_a)^2} -\frac{\mathcal{B}_a^{(0,-1)}}{(1-z_a)}\right]\nonumber \\
&+\mathcal{B}_a^{(0,-2)} \left[-\frac{2}{\varepsilon} +\frac{1}{x_a} -\frac{1}{1-x_a} + 2 \log \left(\frac{1-x_a}{x_a} \right) \right] \nonumber \\
&+\mathcal{B}_a^{(0,-1)}\left[ -\frac{1}{\varepsilon} +\frac{1-x_a}{x_a} + \log \left( \frac{1-x_a}{x_a}\right)\right].
\label{eq:beamplusprescr}
\end{align}
We have included the $z_a^{-2+\varepsilon}$ factor from the phase
space of Eq.~(\ref{eq:PSbeama}).  The beam $a$ contribution to the cross section at NLP-LL  is
\begin{align}
\frac{\d\sigma^\text{NLP,LL}_\text{beam $a$}}{\d Q^2\, \d Y\, \d p_T\, \d \eta}=&\frac{\d\hat{\text{PS}}_\text{Born}}{\d Q^2\, \d Y\, \d p_T\, \d \eta} \left( \frac{\alpha_s}{4 \pi}\right)\frac{\left( 4 \pi \mu_0^2\right)^\varepsilon}{\Gamma\left( 1-\varepsilon\right)} \left(\sqrt{s} x_a \rho_a \right)^{1-\varepsilon}\Tau^{-\varepsilon} \left(-\frac{1}{\varepsilon} \right)\Bigg\{2 \mathcal{B}_a^{(0,-2)} + \mathcal{B}_a^{(0,-1)}\Bigg\}.
\end{align}
%
\subsection{Jet region}
\label{sec:NLPjet}

The jet region treatment proceeds analogously to the beam region.  We define the product of matrix element and phase space, integrated over the azimuthal angle
\begin{equation}
\mathcal{J} \left( \Tau,z_J\right) \equiv\left(\frac{1}{4\pi \alpha_s} \right)\frac{\Omega_{d-3}}{\Omega_{d-2}} \int_0^\pi \d \phi \left( \sin^2\phi\right)^{-\varepsilon}\mathcal{M}_\text{jet} \left(\Tau,z_J,\phi \right) \Phi_\text{jet} \left( \Tau, z_J, \phi\right).
\end{equation}
We write the integral in $z_J$ as
\begin{align}
\int_{0}^\frac12  \d z_J \, z_J^{-\varepsilon} \left(1-z_J \right)^{-\varepsilon}  \mathcal{J}^{(0)}(z_J) &=\int_{0}^\frac12   \d z_J \left[\mathcal{J}^{(0)}(z_J) -\frac{\mathcal{J}^{(0,-2)}}{z_J^2} -\frac{\mathcal{J}^{(0,-1)}}{z_J}\right]\nonumber \\
&-\mathcal{J}^{(0,-2)} +\left(-\frac{1}{\varepsilon} -\log2\right)\mathcal{J}^{(0,-1)}.
\label{eq:jetplusprescr}
\end{align}
The cross section at NLP-LL is
\begin{align}
\frac{\d\sigma^\text{NLP,LL}_\text{jet}}{\d Q^2\, \d Y\, \d p_T\, \d \eta}=&\frac{\d\hat{\text{PS}}_\text{Born}}{\d Q^2\, \d Y\, \d p_T\, \d \eta} \left( \frac{\alpha_s}{4 \pi}\right)\frac{\left( 4 \pi \mu_0^2\right)^\varepsilon}{\Gamma\left( 1-\varepsilon\right)} \left(2 p_T \rho_J \cosh \eta \right)^{1-\varepsilon}\Tau^{-\varepsilon} \left(-\frac{1}{\varepsilon} \right)\Bigg\{ \mathcal{J}^{(0,-1)}\Bigg\}.
\end{align}
%
\subsection{Cancellation of poles}
A strong consistency check of our computation is the cancellation of $\varepsilon^{-1}$ poles. Poles come from the soft function, the two beam functions and the jet function:
\begin{align}
\frac{\d\sigma^\text{NLP,pole}_\text{soft}}{\d Q^2\, \d Y\, \d p_T\, \d \eta} &=\frac{\d \hat{\text{PS}}_\text{Born}}{\d Q^2\, \d Y\, \d p_T\, \d \eta} \left( \frac{\alpha_s}{4 \pi}\right)\left(\frac{1}{\varepsilon}\right)\Bigg\{ 
 \frac{\mathcal{S}^{(-1,0,0)}_{2J} +\mathcal{S}^{(0,-1,0)}_{2J}}{  \hat{s}_{ab}}+ \frac{\mathcal{S}^{(-1,0,0)}_{1J} +\mathcal{S}^{(0,0,-1)}_{2J}}{   \hat{s}_{aJ}} \nonumber \\
&+ \frac{\mathcal{S}^{(0,-1,0)}_{1J} +\mathcal{S}^{(0,0,-1)}_{2J}}{   \hat{s}_{bJ} } 
+\frac{\mathcal{S}_{2J}^{(-1,-1,1)} \hat{s}_{bJ}+\mathcal{S}_{2J}^{(-1,-1,1)} \hat{s}_{aJ}}{  \hat{s}_{ab}^2}\nonumber \\
& +\frac{\mathcal{S}_{2J}^{(-1,1,-1)} \hat{s}_{bJ}+\mathcal{S}_{1J}^{(-1,1,-1)} \hat{s}_{ab}}{ \hat{s}_{aJ}^2} 
+\frac{\mathcal{S}_{2J}^{(1,-1,-1)} \hat{s}_{aJ}+\mathcal{S}_{1J}^{(1,-1,-1)} \hat{s}_{ab}}{ \hat{s}_{bJ}^2}\Bigg\}, \\
\frac{\d\sigma^\text{NLP,pole}_\text{beam $a$}}{\d Q^2\, \d Y\, \d p_T\, \d \eta}&=\frac{\d\hat{\text{PS}}_\text{Born}}{\d Q^2\, \d Y\, \d p_T\, \d \eta} \left( \frac{\alpha_s}{4 \pi}\right) \left(\sqrt{s} x_a \rho_a \right) \left(-\frac{1}{\varepsilon} \right)\Bigg\{2 \mathcal{B}_a^{(0,-2)} + \mathcal{B}_a^{(0,-1)}\Bigg\},\\
\frac{\d\sigma^\text{NLP,pole}_\text{beam $b$}}{\d Q^2\, \d Y\, \d p_T\, \d \eta}&=\frac{\d\hat{\text{PS}}_\text{Born}}{\d Q^2\, \d Y\, \d p_T\, \d \eta} \left( \frac{\alpha_s}{4 \pi}\right) \left(\sqrt{s} x_b \rho_b \right) \left(-\frac{1}{\varepsilon} \right)\Bigg\{2 \mathcal{B}_b^{(0,-2)} + \mathcal{B}_b^{(0,-1)}\Bigg\},\\
\frac{\d\sigma^\text{NLP,pole}_\text{jet}}{\d Q^2\, \d Y\, \d p_T\, \d \eta}&=\frac{\d\hat{\text{PS}}_\text{Born}}{\d Q^2\, \d Y\, \d p_T\, \d \eta} \left( \frac{\alpha_s}{4 \pi}\right) 2 p_T \rho_J \cosh \eta  \left(-\frac{1}{\varepsilon} \right)\Bigg\{ \mathcal{J}^{(0,-1)}\Bigg\}.
\end{align}
Thanks to the relations between the beam and jet matrix element
expansion coefficients and the soft matrix element expansion
coefficients that we derived in Section~\ref{sec:ME}, together with
the phase space expansion coefficients, the
following relations required for pole cancellation are indeed satisfied:
\begin{align}
&\frac{\mathcal{S}^{(-1,0,0)}_{2J} +\mathcal{S}^{(0,-1,0)}_{2J}}{  \hat{s}_{ab}}+ \frac{\mathcal{S}^{(0,0,-1)}_{2J}}{   \hat{s}_{aJ}}
+ \frac{\mathcal{S}^{(0,0,-1)}_{2J}}{   \hat{s}_{bJ} } 
+\frac{\mathcal{S}_{2J}^{(-1,-1,1)} \hat{s}_{bJ}+\mathcal{S}_{2J}^{(-1,-1,1)} \hat{s}_{aJ}}{  \hat{s}_{ab}^2} +\frac{\mathcal{S}_{2J}^{(-1,1,-1)} \hat{s}_{bJ}}{ \hat{s}_{aJ}^2} \nonumber \\
&+\frac{\mathcal{S}_{2J}^{(1,-1,-1)} \hat{s}_{aJ}}{ \hat{s}_{bJ}^2}=
 \left(\sqrt{s} x_a \rho_a \right) \left[ 2 \mathcal{B}_a^{(0,-2)} + \mathcal{B}_a^{(0,-1)}\right]
 +\left(\sqrt{s} x_b \rho_b \right) \left[ 2 \mathcal{B}_b^{(0,-2)} + \mathcal{B}_b^{(0,-1)}\right],
\end{align}
\begin{align}
 \frac{\mathcal{S}^{(-1,0,0)}_{1J} }{   \hat{s}_{aJ}}+ \frac{\mathcal{S}^{(0,-1,0)}_{1J} }{   \hat{s}_{bJ} } 
 +\frac{\mathcal{S}_{1J}^{(-1,1,-1)} \hat{s}_{ab}}{ \hat{s}_{aJ}^2} 
+\frac{\mathcal{S}_{1J}^{(1,-1,-1)} \hat{s}_{ab}}{ \hat{s}_{bJ}^2}=(2 p_T \rho_J \cosh \eta) \mathcal{J}^{(0,-1)}.
\end{align}
We note that these consistency relations are satisfied separately for
each term in the integrand of Eq.~\eqref{eq:Sfunctions}, and also that
the contribution of the non-hemisphere poles is crucial to obtaining
this cancellation.
\subsection{Summary of the NLP-LL result}
This section contains the main result of our paper. We already
anticipated the final form of the NLP cross section in
Eq.~\eqref{eq:NLPXS}. By
expanding in $\varepsilon$, we can now write down the coefficients  $Q_\alpha$ and
$C^\text{LL}_\alpha$, and discuss the terms that contribute to $C^\text{NLL}_\alpha$. We start
by listing the logarithm arguments $Q_\alpha$ that naturally appear
when evaluating the cross sections in each region:
\begin{align}
Q_\text{soft $ij$ hemi} &= \mu\sqrt{ \hat{s}_{ij}}, \;\; Q_{\text{soft $ij$ non-hemi}} = \frac{\mu \hat{s}_{im}}{\sqrt{\hat{s}_{ij}}}, \ \\
Q_\text{beam $a$} &=\frac{\mu^2}{\sqrt{s} x_a \rho_a}, \;\; Q_\text{beam $b$} = \frac{\mu^2}{\sqrt{s} x_b \rho_b},\\
Q_\text{jet} &=\frac{\mu^2}{2 p_T \rho_J \cosh \eta}.
\label{eq:Qs}
\end{align}
We note that these logarithmic arguments can be changed by changing the choice of $\rho_i$, which shifts
terms between the LL and NLL contributions.  We now list LL coefficients:
\begin{align}
C^\text{LL}_\text{soft $ab$ hemi} &= -\frac{\mathcal{S}_{2J}^{(0,-1,0)}}{\hat{s}_{ab}} -\frac{2 \mathcal{S}_{2J}^{(-1,-1,1)}\hat{s}_{bJ}}{\hat{s}_{ab}^2}, 
\qquad &C^\text{LL}_\text{soft $ba$ hemi}&= -\frac{\mathcal{S}_{2J}^{(-1,0,0)}}{\hat{s}_{ab}}-\frac{2 \mathcal{S}_{2J}^{(-1,-1,1)}\hat{s}_{aJ}}{\hat{s}_{ab}^2},\label{eq:CLLbegin}\\
C^\text{LL}_\text{soft $aJ$ hemi} &= -\frac{\mathcal{S}_{2J}^{(0,0,-1)}}{\hat{s}_{aJ}} -\frac{2 \mathcal{S}_{2J}^{(-1,1,-1)}\hat{s}_{bJ}}{\hat{s}_{aJ}^2} ,
\qquad &C^\text{LL}_\text{soft $Ja$ hemi}&= -\frac{\mathcal{S}_{1J}^{(-1,0,0)}}{\hat{s}_{aJ}}-\frac{2 \mathcal{S}_{1J}^{(-1,1,-1)}\hat{s}_{ab}}{\hat{s}_{aJ}^2},\\
C^\text{LL}_\text{soft $bJ$ hemi} &= -\frac{\mathcal{S}_{2J}^{(0,0,-1)}}{\hat{s}_{bJ}} -\frac{2 \mathcal{S}_{2J}^{(1,-1,-1)}\hat{s}_{aJ}}{\hat{s}_{bJ}^2} ,
\qquad &C^\text{LL}_\text{soft $Jb$ hemi}&= -\frac{\mathcal{S}_{1J}^{(0,-1,0)}}{\hat{s}_{bJ}}-\frac{2 \mathcal{S}_{1J}^{(1,-1,-1)}\hat{s}_{ab}}{\hat{s}_{bJ}^2},
\end{align}
\begin{align}
C^\text{LL}_\text{soft $ab$ non-hemi} &= -\left( \frac{\mathcal{S}_{1J}^{(-1,0,0)}}{\hat{s}_{aJ}}+\frac{\mathcal{S}_{1J}^{(0,-1,0)}}{\hat{s}_{bJ}}\right) \Theta \left(\hat{s}_{bJ} - \hat{s}_{aJ} \right),\\
C^\text{LL}_\text{soft $ba$ non-hemi}&= -\left( \frac{\mathcal{S}_{1J}^{(-1,0,0)}}{\hat{s}_{aJ}}+\frac{\mathcal{S}_{1J}^{(0,-1,0)}}{\hat{s}_{bJ}}\right) \Theta \left(\hat{s}_{aJ} - \hat{s}_{bJ} \right),\\
C^\text{LL}_\text{soft $aJ$ non-hemi} &= -\left( \frac{\mathcal{S}_{2J}^{(-1,0,0)}}{\hat{s}_{ab}}+\frac{\mathcal{S}_{2J}^{(0,0,-1)}}{\hat{s}_{bJ}}\right) \Theta \left(\hat{s}_{bJ} - \hat{s}_{ab} \right),\\
C^\text{LL}_\text{soft $Ja$ non-hemi}&= -\left( \frac{\mathcal{S}_{2J}^{(-1,0,0)}}{\hat{s}_{ab}}+\frac{\mathcal{S}_{2J}^{(0,0,-1)}}{\hat{s}_{bJ}}\right) \Theta \left(\hat{s}_{ab} - \hat{s}_{bJ} \right),\\
C^\text{LL}_\text{soft $bJ$ non-hemi} &= -\left( \frac{\mathcal{S}_{2J}^{(0,-1,0)}}{\hat{s}_{ab}}+\frac{\mathcal{S}_{2J}^{(0,0,-1)}}{\hat{s}_{aJ}}\right) \Theta \left(\hat{s}_{aJ} - \hat{s}_{ab} \right) ,\\
C^\text{LL}_\text{soft $Jb$ non-hemi}&= -\left( \frac{\mathcal{S}_{2J}^{(0,-1,0)}}{\hat{s}_{ab}}+\frac{\mathcal{S}_{2J}^{(0,0,-1)}}{\hat{s}_{aJ}}\right) \Theta \left(\hat{s}_{ab} - \hat{s}_{aJ} \right),
\end{align}
\begin{align}
C_\text{beam $a$}^\text{LL}&=  \sqrt{s} x_a \rho_a \left[2\mathcal{B}_a^{(0,-2)} + \mathcal{B}_a^{(0,-1)}\right], \\
C_\text{beam $b$}^\text{LL}&=  \sqrt{s} x_b \rho_b \left[2\mathcal{B}_b^{(0,-2)} + \mathcal{B}_b^{(0,-1)}\right],  \\
C_\text{jet}^\text{LL}&=  2 p_T \rho_J \cosh \eta\,  \mathcal{J}^{(0,-1)} .\label{eq:CLLend}
\end{align}
We recall that the $\mathcal{S}_{nJ}$ coefficients are defined in
Sec.~\ref{sec:NLPsoft}, while the $\mathcal{B}_i$ and $\mathcal{J}_a$
coefficients are defined respectively in Sections ~\ref{sec:NLPbeam}
and~\ref{sec:NLPjet}.  All three structures can be written in terms of
process-independent phase-space corrections given in the Appendix.  From Sec.~\ref{sec:ME} we see that the matrix elements
appearing in these structures can be expressed in terms of the
universal next-to-soft matrix element expansion.  This demonstrates
that the NLP-LL cross section can be written in terms of universal
factors valid for any 1-jet process. We provide an explicit expression for the coefficients~(\ref{eq:CLLbegin}--\ref{eq:CLLend}) in a supplemental file.
\subsection{NLP-NLL contributions}
\label{sec:NLPNLL}
In this section we analyze the terms that contribute to the NLP-NLL
cross section. In the color-singlet case, it was
observed in Refs.~\cite{Boughezal:2018mvf,Ebert:2018lzn} that different
definitions of N-jettiness, corresponding to different values for
$\rho_a$ and $\rho_b$ in Eq.~\eqref{eq:taudef}, can produce very different power corrections. In particular, the hadronic definition ($\rho_a = \rho_b = 1$) had much larger power corrections than the leptonic definition ($\rho_a = \sqrt{s} x_b$, $\rho_b=\sqrt{s} x_a$). It was found that LL corrections alone provide a sufficient improvement to the N-jettiness cross section in the leptonic case, whereas NLL corrections are necessary in the hadronic case.

For processes with one jet in the final state, the NLL power
correction are inherently process dependent since they require the subleading collinear matrix elements.  It is not unreasonable to assume that, like in the color singlet case, there is a choice of $\rho_i$ that reduces the impact of power corrections, avoiding the need to implement NLL contributions.  We therefore do not provide a complete analytical computation of the NLL contribution, and only outline how such contributions arise.
\begin{itemize}
\item Soft region: starting from Eq.~\eqref{eq:Sfunctions}, we write down the measurement function explicitly. Then, we expand in $\varepsilon$ and consider the finite contributions. The hemisphere contributions are straightforward:
\begin{align}
\frac{\d\sigma^\text{NLP,NLL}_\text{soft hemi}}{\d Q^2\, \d Y\, \d p_T\, \d \eta} =&\frac{\d \hat{\text{PS}}_\text{Born}}{\d Q^2\, \d Y\, \d p_T\, \d \eta} \left( \frac{\alpha_s}{4 \pi}\right)\Bigg\{ 
-\frac{\mathcal{S}^{(-1,0,0)}_{2J} +\mathcal{S}^{(0,-1,0)}_{2J}}{2  \hat{s}_{ab}}
- \frac{\mathcal{S}^{(-1,0,0)}_{2J} +\mathcal{S}^{(0,0,-1)}_{1J}}{2  \hat{s}_{aJ}} \nonumber \\
&-\frac{\mathcal{S}^{(0,-1,0)}_{2J} +\mathcal{S}^{(0,0,-1)}_{1J}}{2  \hat{s}_{bJ}} 
-\frac{\mathcal{S}_{2J}^{(-1,-1,1)} \hat{s}_{aJ}+\mathcal{S}_{2J}^{(-1,-1,1)} \hat{s}_{bJ}}{\left( \hat{s}_{ab}\right)^{2}}\nonumber \\
&-\frac{\mathcal{S}_{2J}^{(-1,1,-1)} \hat{s}_{ab}+\mathcal{S}_{1J}^{(-1,1,-1)} \hat{s}_{bJ}}{\left( \hat{s}_{aJ}\right)^{2}} 
-\frac{\mathcal{S}_{2J}^{(1,-1,-1)} \hat{s}_{ab}+\mathcal{S}_{1J}^{(1,-1,-1)} \hat{s}_{aJ}}{\left( \hat{s}_{bJ}\right)^{2}}\Bigg\}.
\end{align}
As for the non-hemisphere contributions, we identified three different
terms in the $ij,m$ non-hemisphere region in
Eq.~\eqref{eq:nonhemi123}. The first and second term will produce
finite contributions that can be read from Eq.~\eqref{eq:gintegral}:
\begin{align}
\int_{x_\text{min}}^{x_\text{max}} \d x \, \frac{g(x)}{ \left|x-x_0 \right|^{1-2\varepsilon}} &=\text{pole}+\int_{x_\text{min}}^{x_\text{max}} \d x \frac{g(x) -g(x_0)}{\left|x-x_0\right|}+g(x_0) \, K_\text{non-hemi} \left(x_0,x_\text{min},x_\text{max} \right),
 \end{align}
while the third term is already finite. The NLL contribution to the cross section will be the sum of the finite parts of the three $ij,m$ terms, plus the sum of all $ij,i$ terms. 
\item Beam region: the NLL contributions come from the finite terms in Eq.~\eqref{eq:beamplusprescr}, plus contributions coming from the $\Tau$ expansion of the lower integration limit in $z_a$ in Eq.~\eqref{eq:zalowerlimit}.
\item Jet region: similar to the beam region, there are finite NLL contributions that can be obtained from Eq.~\eqref{eq:jetplusprescr}.  There are also contributions from the  $\Tau$ expansion of the upper integration limit in $z_J$ in Eq.~\eqref{eq:zJlimit}.
\end{itemize}
%
\section{Numerics}
\label{sec:numerics}
In this section we provide a numerical validation of our analytic
results. We consider the partonic process $q\bar{q}\to Z +g$ at $\sqrt{s}=14$
TeV. We use the CT10 PDF set~\cite{Gao:2013xoa} with fixed scales $\mu_R=\mu_F=m_Z$, and
we choose $\rho_a = \rho_b = \rho_J = 1$. In order to study the
behavior of the power corrections, we assume that the N-jettiness
cross section for a very small value of $\Tauc$ (0.0001 GeV) is a good
approximation of the exact NLO cross section. For the channel considered, we have checked that the difference between the two is about 0.5\%.  We then study the
difference between this low-$\Tauc$ reference result and the
NLO cross section as a function of $\Tauc$, normalized to the leading order cross section. 
\begin{figure}
\centering
\includegraphics[width=0.8\textwidth]{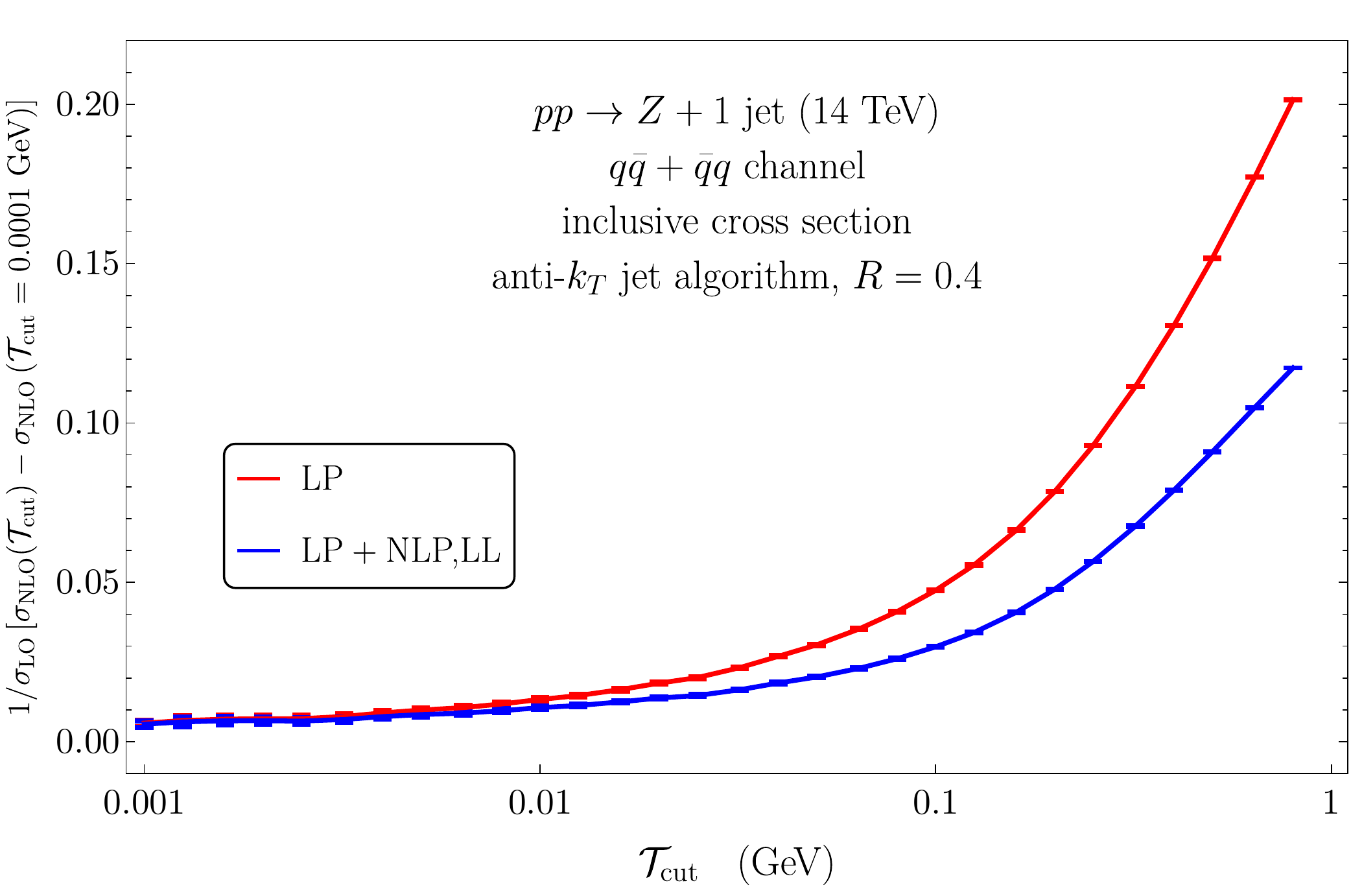}\\
\includegraphics[width=0.8\textwidth]{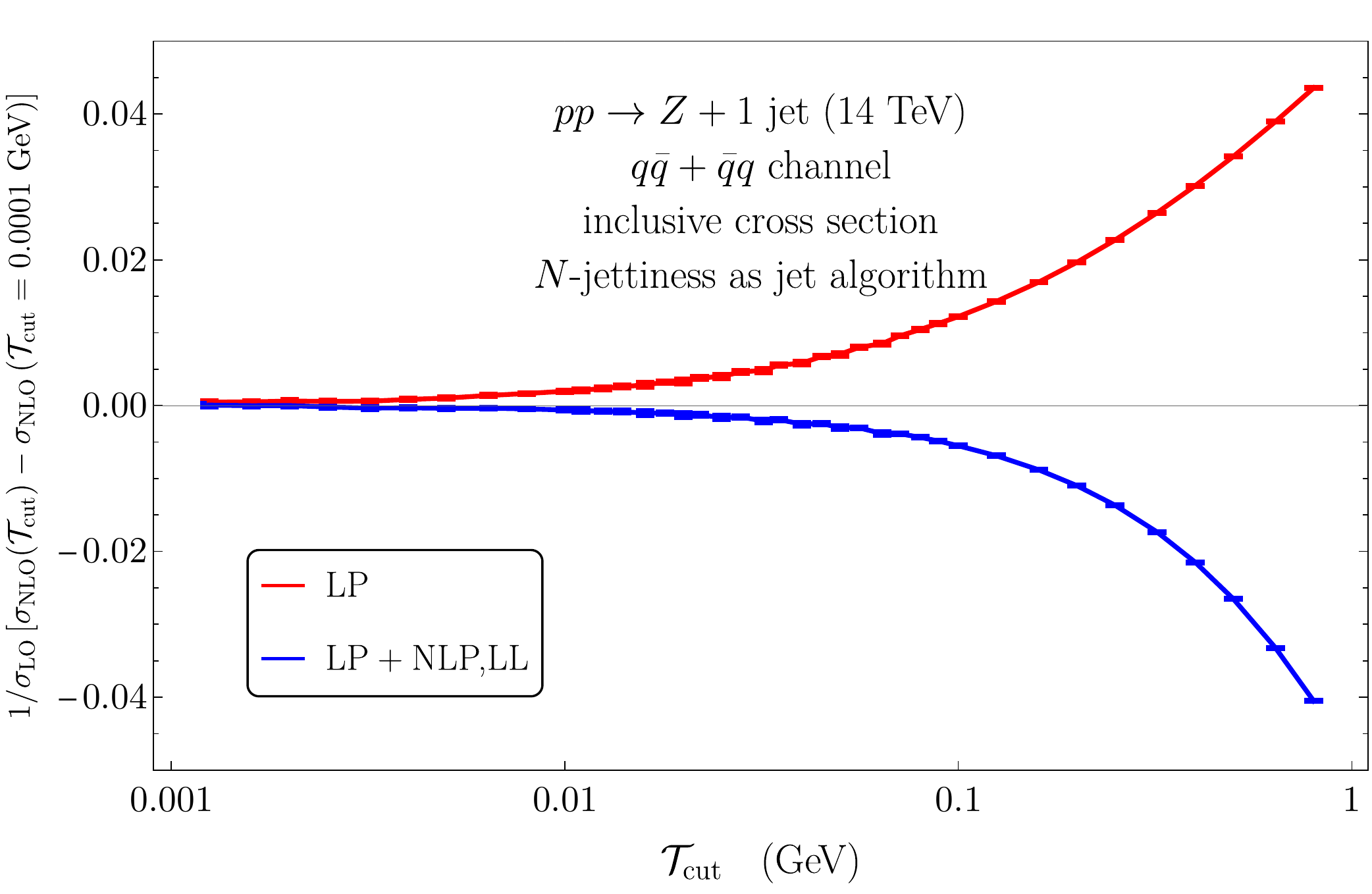}
\caption{Behavior of the NLO cross section as a function of
  $\Tauc$. The red line represents the leading power result, while the
  blue line includes the NLP-LL power corrections. The difference between the top and the bottom insets is the jet algorithm choice for the leading power result. In the top plot, we use MCFM with an anti-$k_T$ pre-clustering jet algorithm. In the bottom plot, we use $N$-jettiness itself as a jet algorithm.}
\label{fig:powcorrplot}
\end{figure}

We first show in Fig.~\ref{fig:powcorrplot} the cross section as a
function of $\Tauc$ when no power corrections are included compared to
when NLP-LL power corrections are included. We obtain the
leading-power cross section in two ways.  We first use MCFM (Monte Carlo for FeMtobarn processes)~\cite{Boughezal:2016wmq} which
implements an anti-$k_T$ pre-clustering algorithm to define
$N$-jettiness. We also use an independent code that treats $N$-jettiness itself as the jet algorithm, according to the framework that we used to compute power corrections in this paper.

We note that the size of the deviation from zero, which includes all power corrections (not just NLP-LL) is significantly larger in the presence of a pre-clustering jet algorithm for all values of $\Tauc$.
We also note that using $N$-jettiness as a jet algorithm and for our choice of $\Tau$ definition, equivalent to the hadronic definition in the color singlet case, the NLP-LL contributions seem to overcorrect the LP result. For the purpose of validating our result, a $\Tau$ definition that produces large power corrections is preferable in order to avoid numerical noise. However, for other applications of the N-jettiness subtraction scheme another definition might be more suitable. 

In order to validate our result for the LL power corrections, we define the full nonsingular cross section as
\begin{equation}
\text{full nonsing.}\left(\Tauc \right) \equiv \frac{\sigma_\text{NLO} \left(\Tauc=0.0001 \text{GeV} \right) -\sigma_\text{NLO} \left(\Tauc\right)}{\sigma_\text{LO}}.
\label{eq:fullnonsing}
\end{equation}
The functional form of the full nonsingular cross section is
\begin{equation}
\text{full nonsing.}\left(\Tauc \right) = A\,  \Tauc \log  \Tauc  + B \, \Tauc + C \, \Tauc^2 \log  \Tauc + D\,  \Tauc^2 + \dots.
\label{eq:fullnonsingfit}
\end{equation}
where the ellipsis denote neglected power corrections at ${\cal
  O}(\Tauc^3)$ and above. We perform a fit to extract the coefficients $A,B,C,D$ and then compare the fitted $A$ with the analytic $A$. For the inclusive cross section we find
\begin{align}
A_\text{fitted}^\text{incl}&=0.0345 \pm 0.0014, \\
A_\text{analytic}^\text{incl}&=0.0346.
\end{align}
This indicates excellent agreement between the fitted and the analytic LL coefficient.
\begin{figure}
\centering
\includegraphics[width=0.85\textwidth]{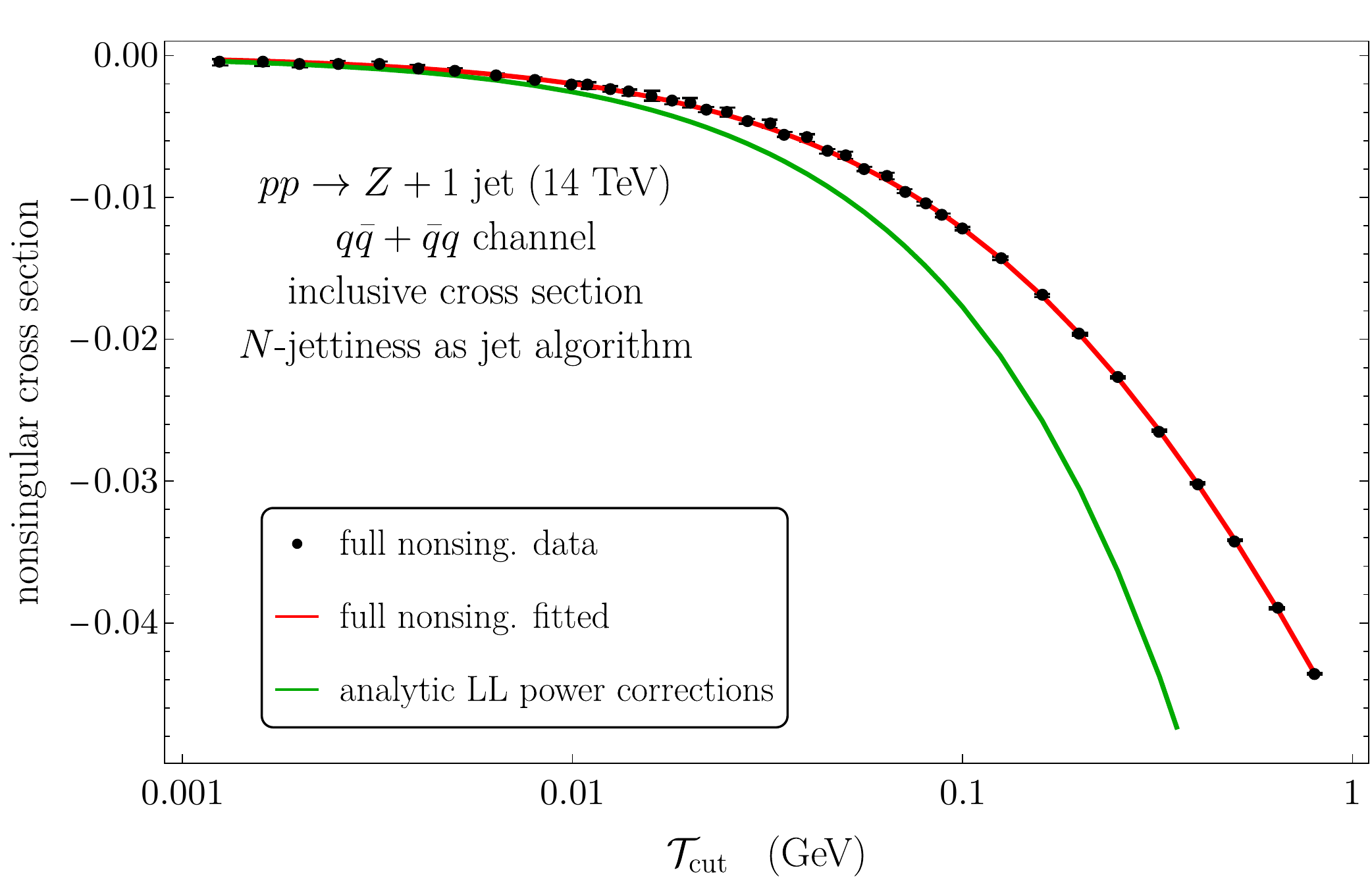}
\caption{Full nonsingular cross section as a function of $\Tauc$ as
  defined in Eq.~\eqref{eq:fullnonsing} for the inclusive case. The
  solid red line represents a fit of the form
  Eq.~\eqref{eq:fullnonsingfit}.  The data refers to the numerical
  results from our code for $Z$+jet production. The solid green line indicates the analytic leading logarithmic power corrections, normalized to the LO cross section.}
\label{fig:nonsingularinclusive}
\end{figure}
In Fig.~\ref{fig:nonsingularinclusive} we plot the full nonsingular cross section as defined in Eq.~\eqref{eq:fullnonsing}, together with the LL power corrections. 

\begin{figure}
\centering
\begin{minipage}{0.48\textwidth}
\centering
\includegraphics[width=0.9\textwidth]{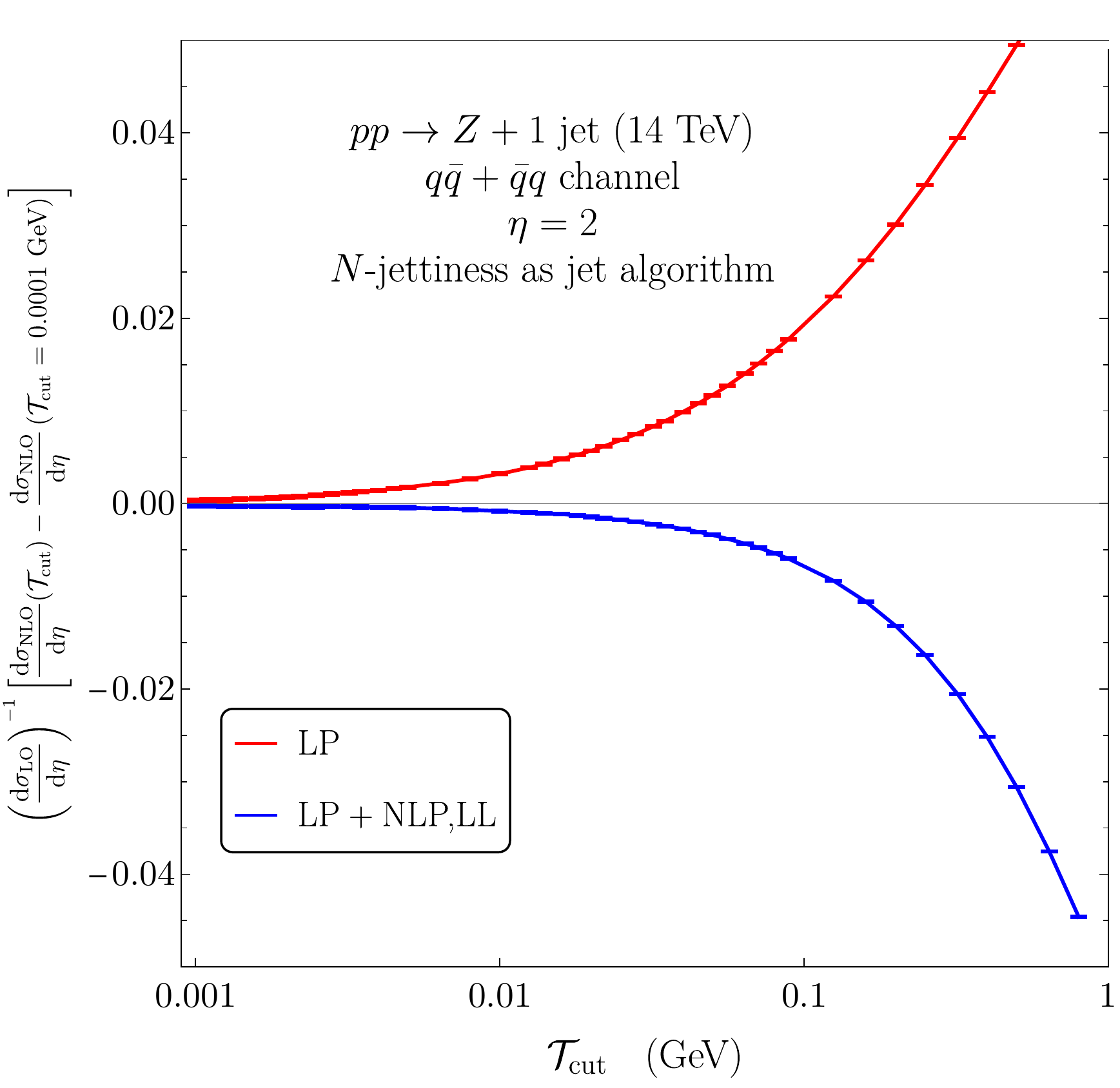} \\
\includegraphics[width=0.9\textwidth]{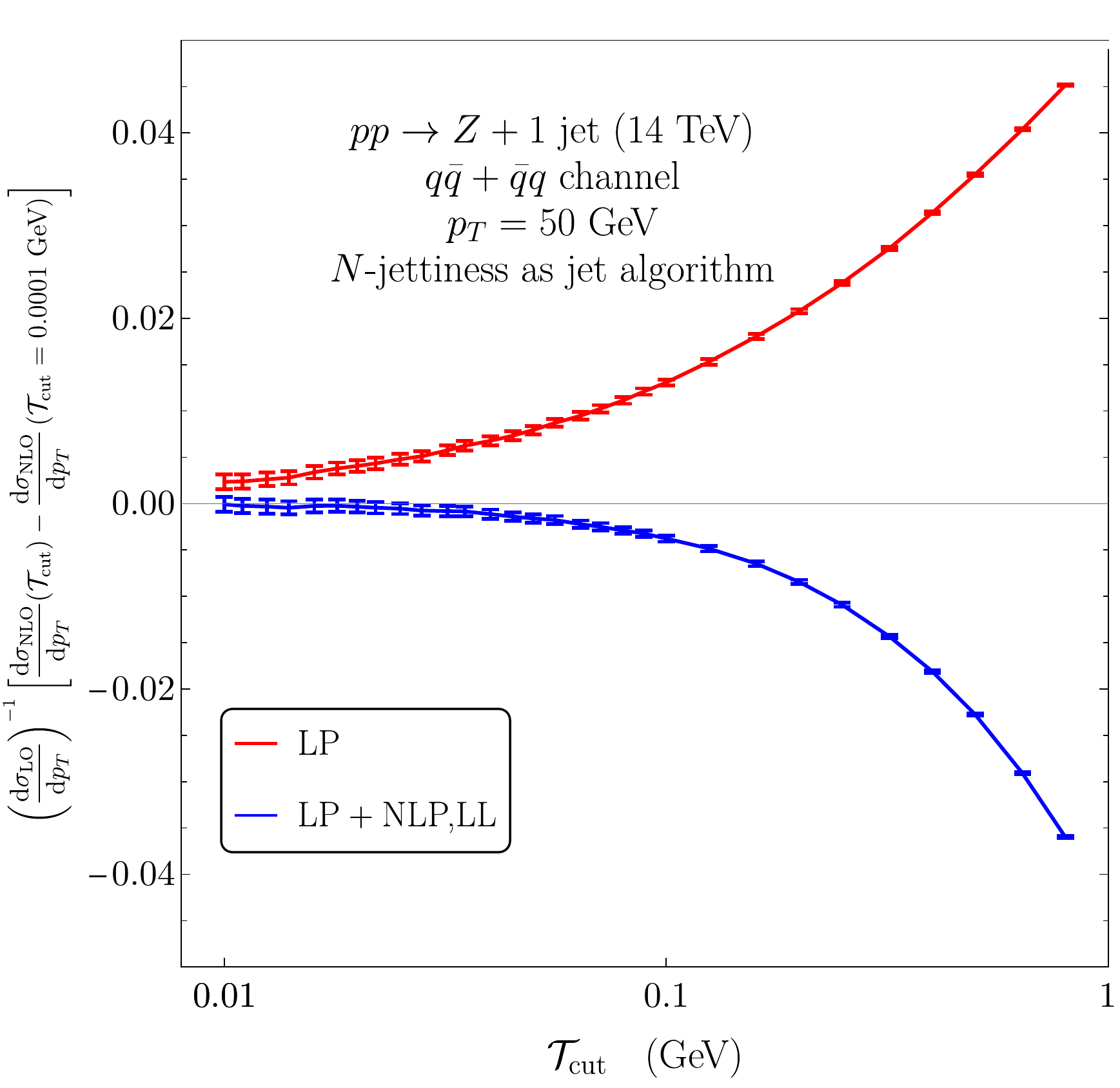} \\
\includegraphics[width=0.9\textwidth]{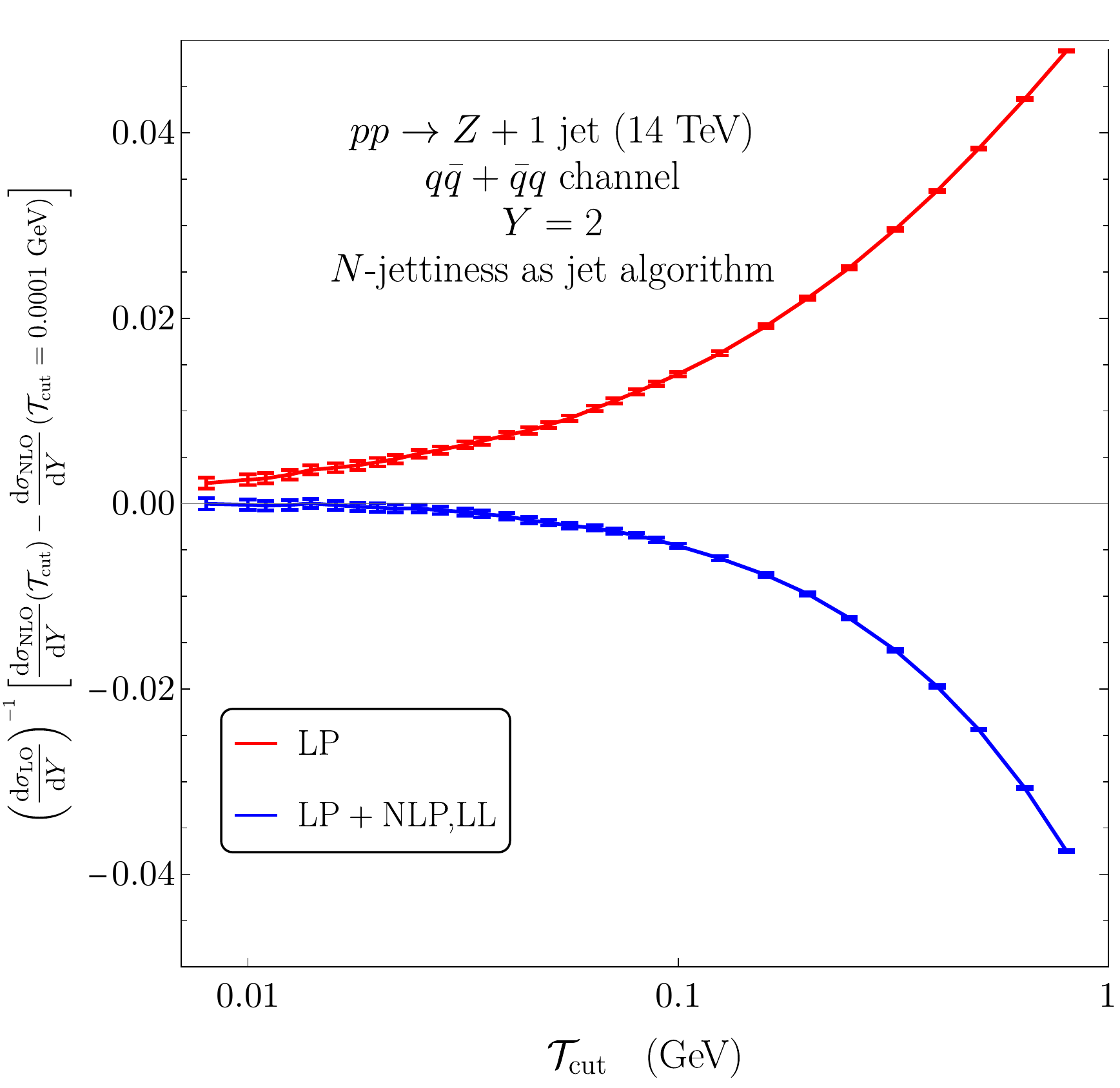} 
\end{minipage}
\begin{minipage}{0.48\textwidth}
\centering
\includegraphics[width=0.9\textwidth]{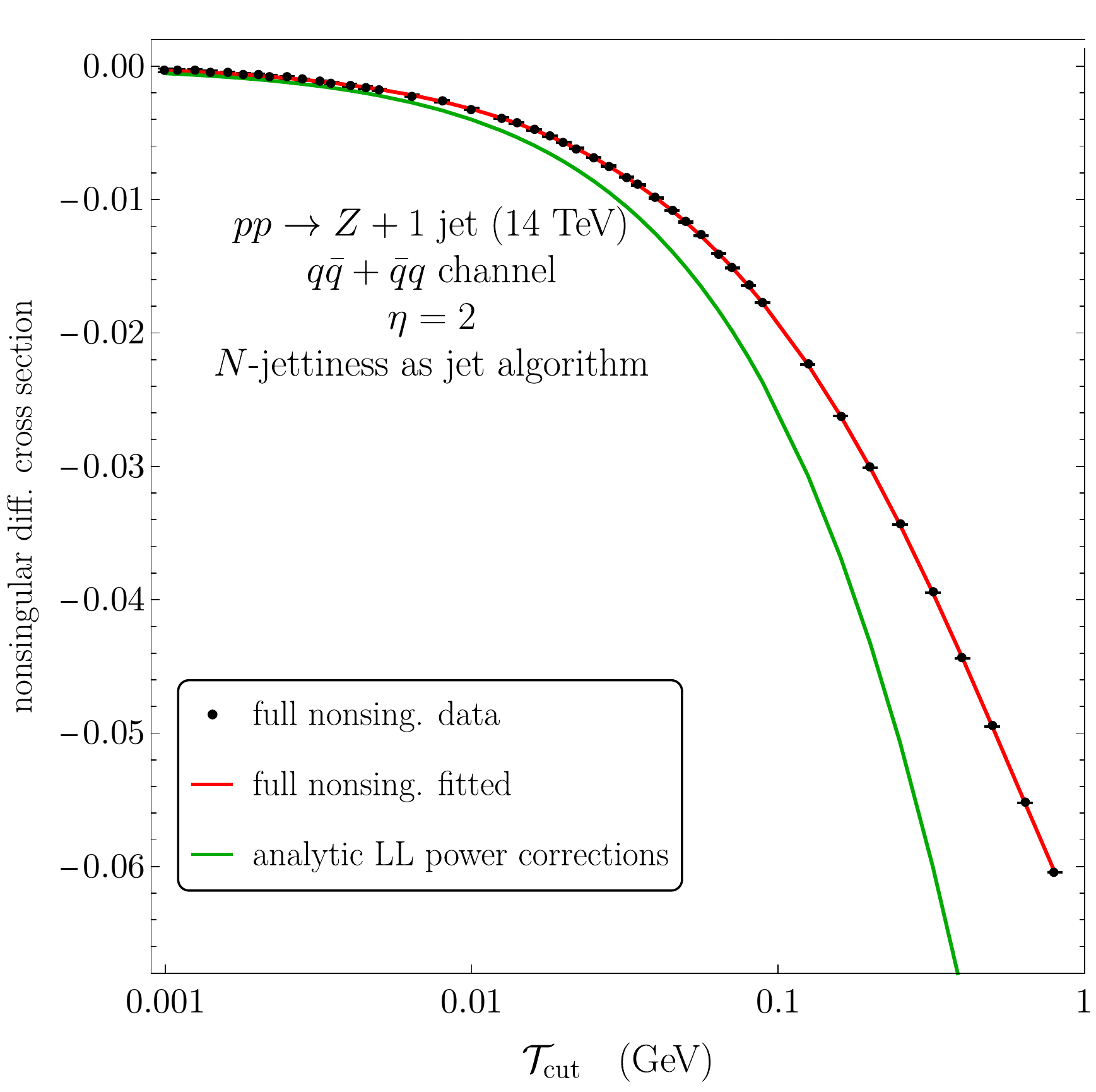}\\
\includegraphics[width=0.9\textwidth]{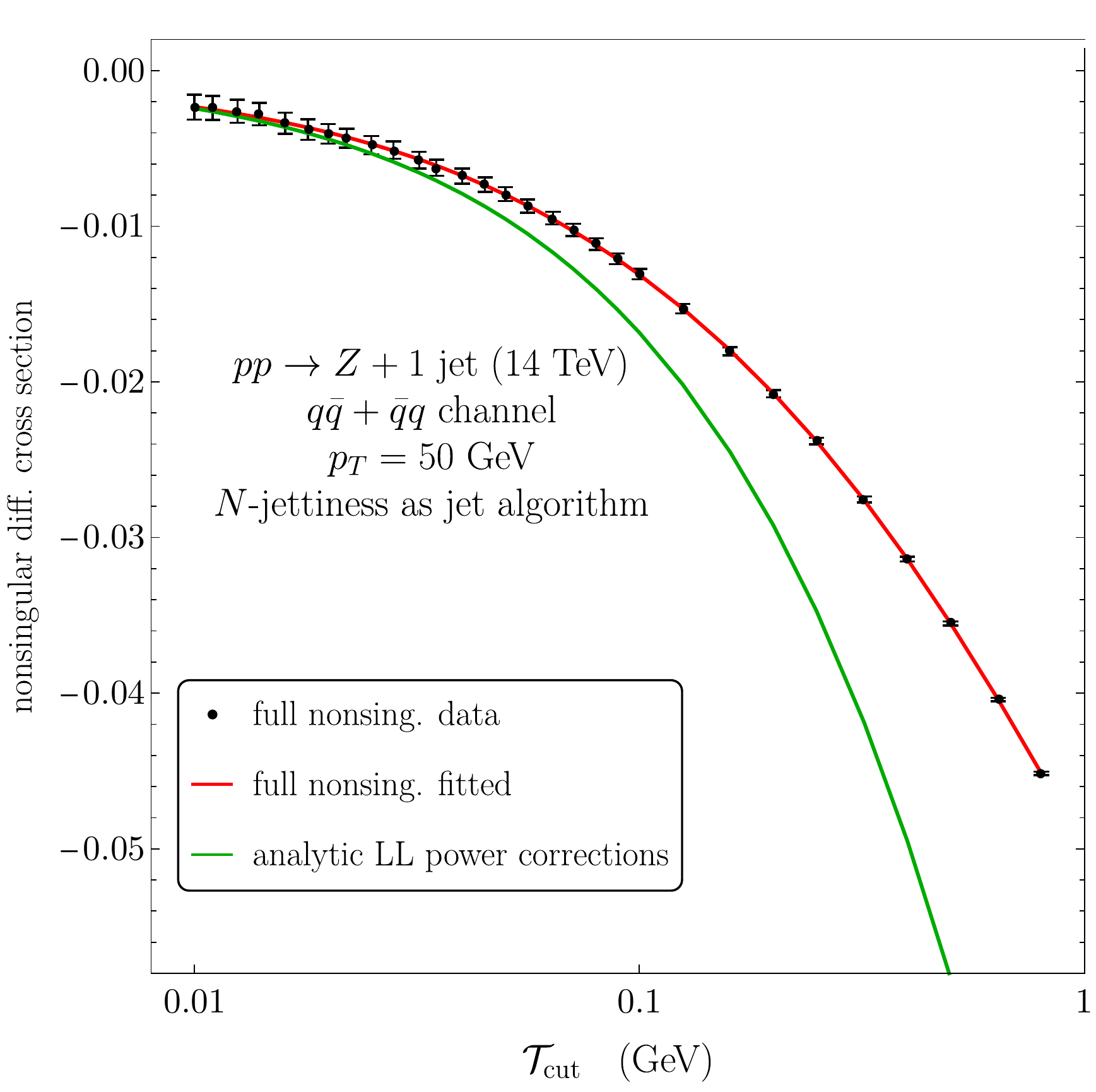} \\
\includegraphics[width=0.9\textwidth]{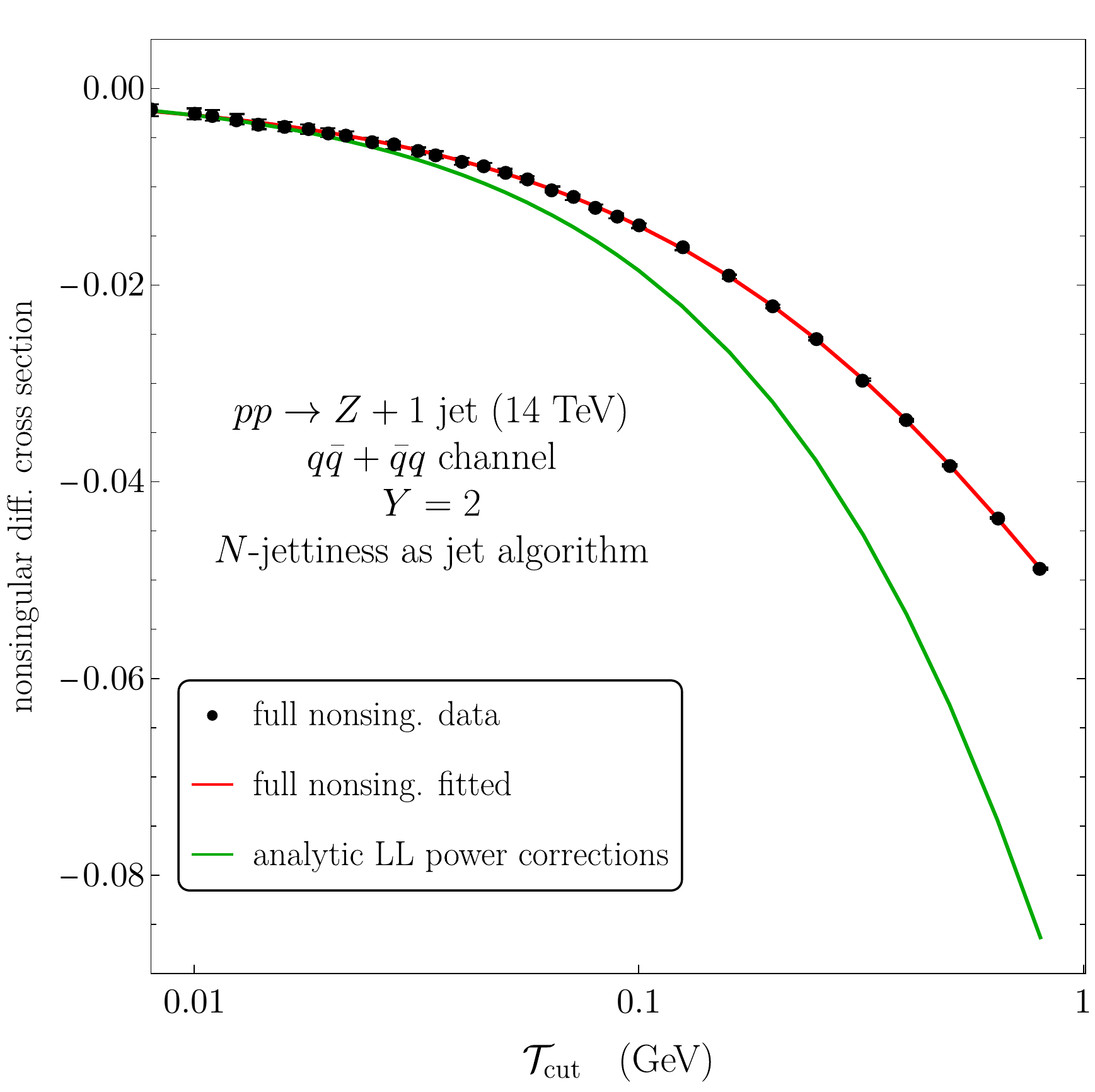}
\end{minipage}
\caption{On the left we show plots analogous to
  Fig.~\ref{fig:powcorrplot}.   On the right we show plots analogous to
  Figure~\ref{fig:nonsingularinclusive} for the differential cross section at the values $\eta=2$ (top), $p_T=50$ GeV (middle) and $Y=2$ (bottom).}
\label{fig:diffplots}
\end{figure}

We have also performed the same validation for the differential cross section in the jet rapidity $\eta$ choosing as a benchmark value $\eta=2$, in the jet transverse momentum $p_T$ choosing as a benchmark value $p_T=50$ GeV and in the vector boson rapidity $Y$ choosing as a benchmark value $Y=2$. The results for the fits and the analytic coefficients are
\begin{align}
A^{\eta=2}_\text{fitted} &=0.0598 \pm 0.0013,& \qquad A^{p_T=50\text{ GeV}}_\text{fitted} &=0.0306 \pm 0.0016,& \qquad A^{Y=2}_\text{fitted} &=0.0321 \pm 0.0046 \nonumber \\
A^{\eta=2}_\text{analytic}&=0.0614,& A^{p_T=50\text{ GeV}}_\text{analytic}&=0.0322,& A^{Y=2}_\text{analytic}&=0.0372
\end{align}
We again find good agreement between our analytic coefficients for the
LL power correction and the fitted results. Plots for the differential cross section are shown in Figure~\ref{fig:diffplots}.

\begin{figure}
\centering
\begin{minipage}{0.48\textwidth}
\centering
\includegraphics[width=\textwidth]{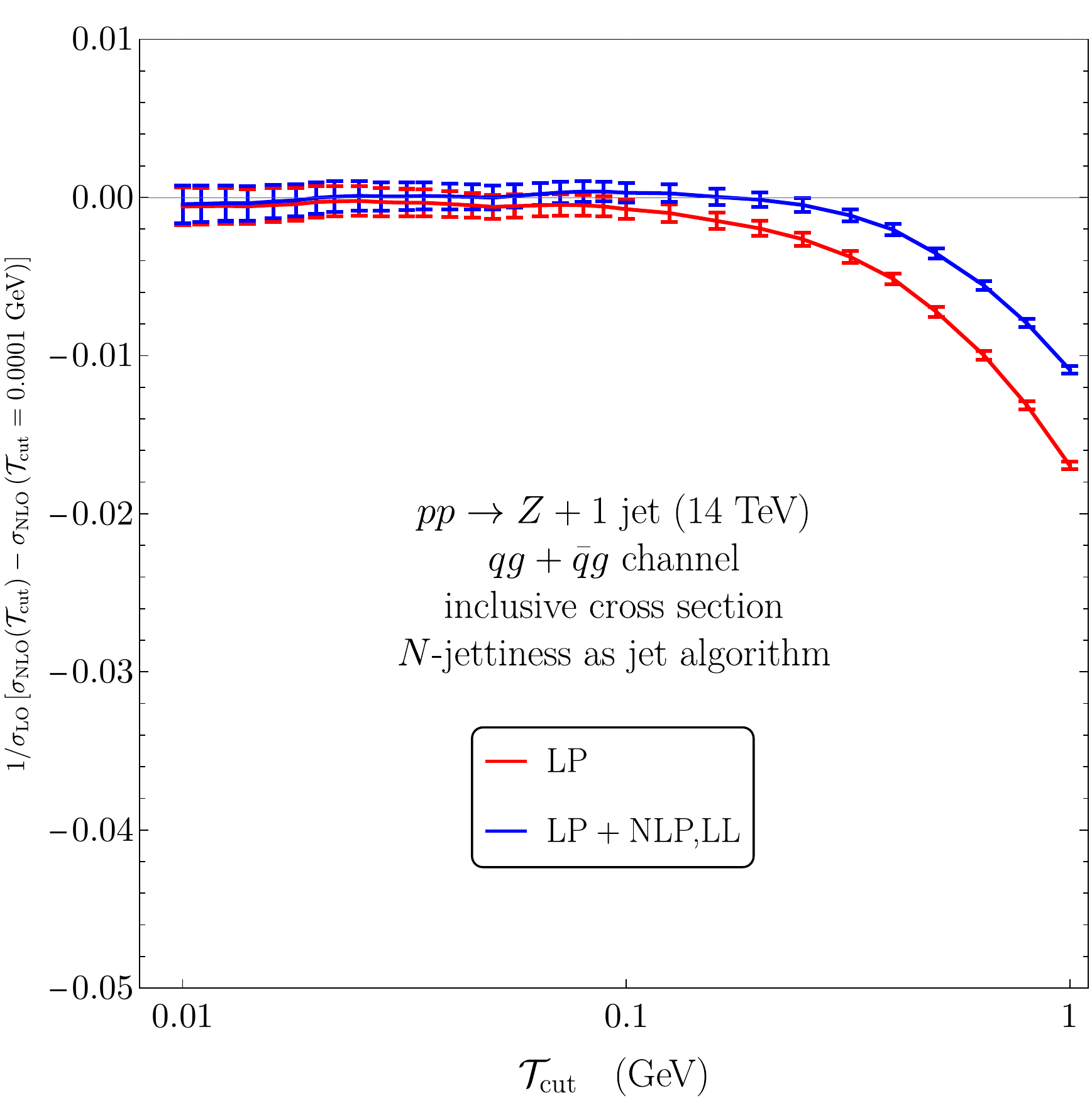} \\
\end{minipage}
\begin{minipage}{0.48\textwidth}
\centering
\includegraphics[width=\textwidth]{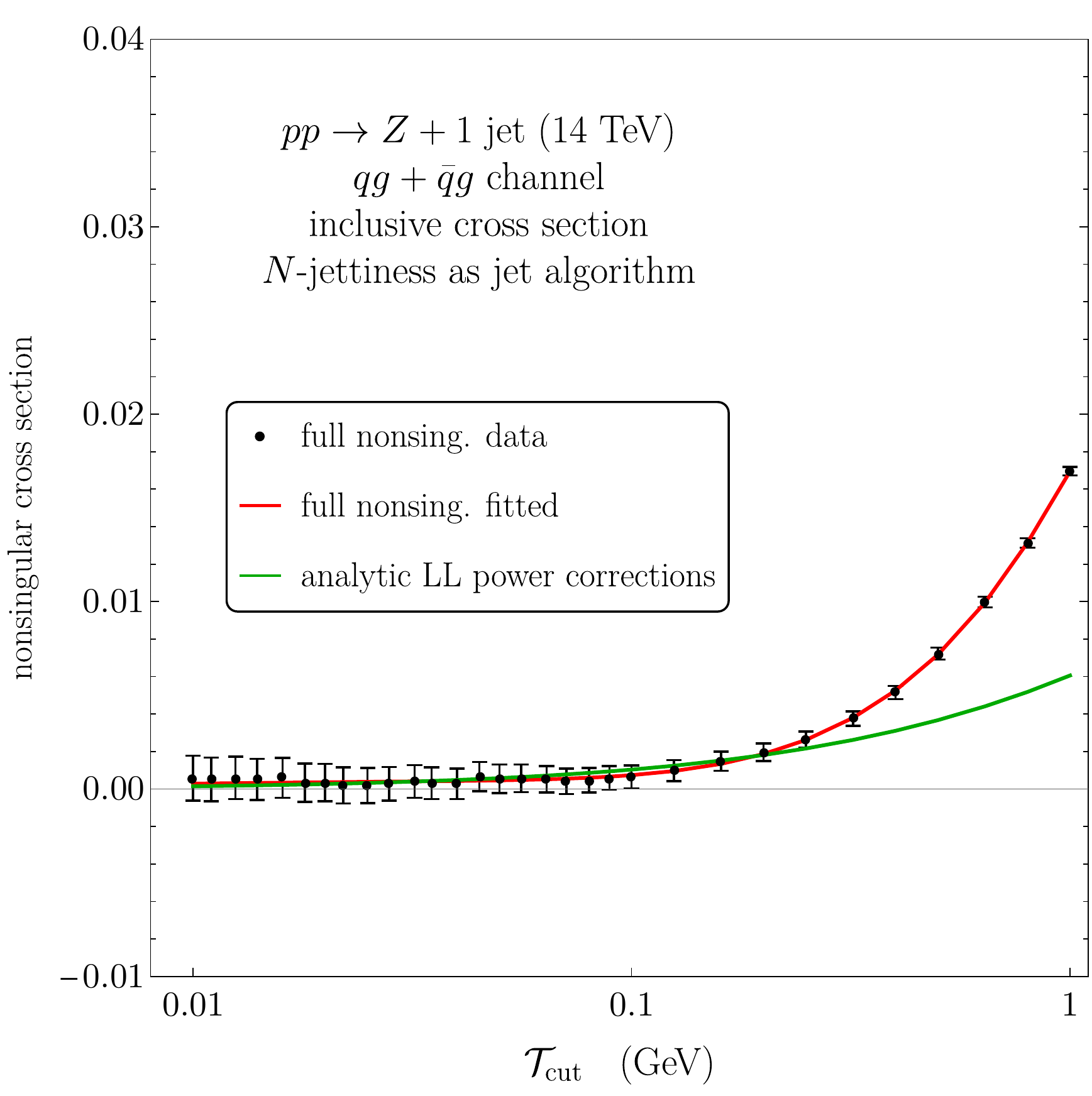}\\
\end{minipage}
\caption{Analogous to Figure~\ref{fig:diffplots} but for the inclusive cross section in the $qg$ channel.}
\label{fig:qgchannel}
\end{figure}

We have also studied the partonic channel $qg \to Z + g$. The LL power corrections for the inclusive cross section are about a factor of 10 smaller than in the $q\bar{q}$ channel in the $\Tauc$ range that determines the fit of the leading log coefficient (between 0.001 and 0.1 GeV). Therefore, the error on the fit is too large to be considered valid. Nevertheless, we provide a plot of the $\Tauc$ dependence of the cross section and the comparison between analytic power corrections and the full nonsingular cross section in Figure~\ref{fig:qgchannel}.

The last numerical study that we performed concerns the impact of
choosing different normalization factors $\rho_i$. As mentioned in
Section~\ref{sec:NLPNLL}, in the case of color singlet it was observed
that the impact of power correction is greatly reduced if one defines $\Tau_0$ in the frame of reference where the color singlet is at rest, rather than in the hadronic frame. Boosting the frame of reference and setting $\rho_i=1$ is equivalent to staying in the hadronic frame and definining the $\rho_i$ factors so as to match what the boosted definition would be. In particular, in the color singlet case it is sufficient to set
\begin{equation}
\rho^\text{lep}_a = e^{-Y} \qquad \qquad \rho_b^\text{lep} = e^Y,
\end{equation}
where $Y$ is the color-singlet rapidity. It was also observed
in~\cite{Campbell:2019gmd} that a similar effect occurs for $V$ + 1
jet processes when one defines $\Tau_1$ in a frame where the system
$V+1$ jet is at rest. We can reproduce this definition with our
framework by boosting to such a frame and determining the $\rho_i$
factors needed to match this definition.  We define the following quantities
\begin{align}
\hat{Q}^2&=\left(p_V+q_J\right)^2= 2 q_a \cdot q_b = s x_a x_b \\
\hat{Y} &= \frac12 \log \left[\frac{\left(p_V+q_J\right) \cdot n_b}{\left(p_V+q_J \right) \cdot n_a} \right]= \frac12 \log  \frac{x_a}{x_b}.
\end{align}
The boosted momenta are then
\begin{equation}
\hat{q}_a^\mu = \frac{\hat{Q}}{2} n_a^\mu, \qquad \qquad \hat{q}_b^\mu = \frac{\hat{Q}}{2} n_b^\mu, \qquad \qquad \hat{q}_J^\mu =\begin{pmatrix} p_T \cosh\left(\eta-\hat{Y}\right) \\ p_T\\ 0 \\p_T \sinh \left(\eta- \hat{Y}\right)\end{pmatrix}.
\end{equation}
As defined in~\cite{Campbell:2019gmd}, the boosted $\Tau$ is
\begin{align}
\Tau_\text{boosted} &= \sum_k \min \left\{\frac{2 \hat{q}_a \cdot \hat{p}_k}{2 \hat{E}_a},\frac{2 \hat{q}_b \cdot \hat{p}_k}{2 \hat{E}_b},\frac{2 \hat{q}_J \cdot \hat{p}_k}{2 \hat{E}_J} \right\}\nonumber \\
&= \sum_k \min \left\{ e^{\hat{Y}} n_a \cdot p_k, e^{-\hat{Y}} n_b \cdot p_k, \frac{\cosh \eta}{\cosh\left(\eta-\hat{Y}\right)} n_J \cdot p_k\right\},
\end{align}
which means that the factors $\rho_i$ for the boosted $\Tau$ definition are
\begin{equation}
\rho_a^\text{boosted} = e^{-\hat{Y}} \qquad \qquad \rho_b^\text{boosted} = e^{\hat{Y}} \qquad \qquad \rho_J^\text{boosted} = \frac{\cosh\left(\eta-\hat{Y}\right)}{\cosh \eta}.
\label{eq:boosteddef}
\end{equation}
A posteriori, having computed the LL power corrections analytically, we can see why this definition has smaller power corrections. In fact, with respect to $\hat{Q}$ and $\hat{Y}$, the initial-state momentum fractions are
\begin{equation}
x_a = \frac{\hat{Q} e^{\hat{Y}}}{\sqrt{s}} \qquad \qquad x_b = \frac{\hat{Q} e^{-\hat{Y}}}{\sqrt{s}}.
\end{equation}
The argument of the logarithms of Eq.~(\eqref{eq:Qs}) in the beam and jet region become
\begin{equation}
Q_\text{beam $a$}^\text{boosted} = Q_\text{beam $b$}^\text{boosted} = \frac{\mu^2}{ \hat{Q}}, \qquad  Q_\text{jet}^\text{boosted} = \frac{\mu^2}{2 p_T \cosh (\eta - \hat{Y})}, 
\end{equation}
clearly smaller than in the hadronic case for large rapidities $\hat{Y}$.

\begin{figure}
\centering
\includegraphics[width=0.8\textwidth]{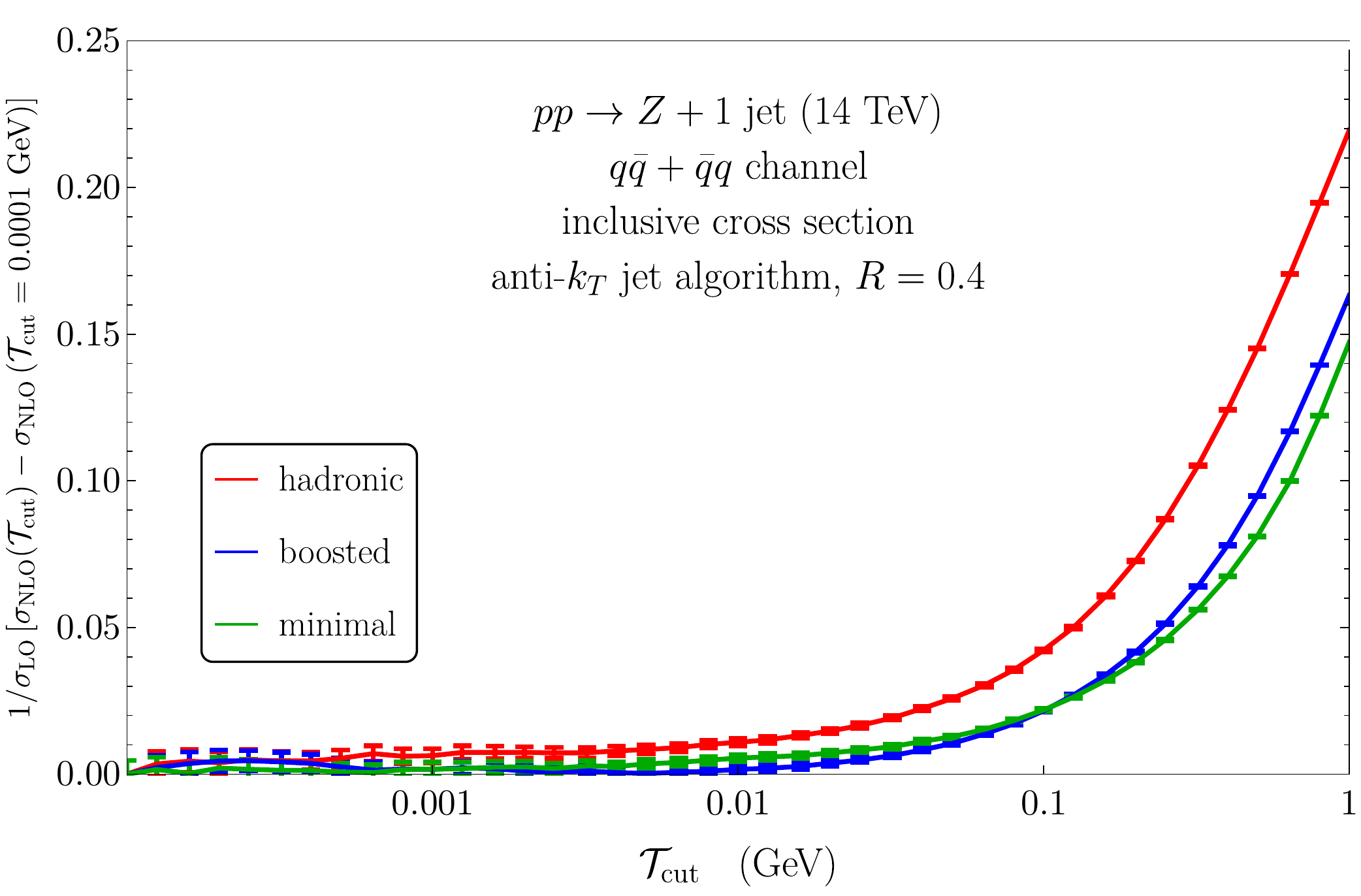}
\caption{$\Tauc$ dependence of the cross section according to three combination of the normalization factors: the hadronic definition ($\rho_a=\rho_b=\rho_J=1$), the boosted definition~\eqref{eq:boosteddef} and the minimal definition~\eqref{eq:minimaldef}.}
\label{fig:taudef}
\end{figure}

We can also define the $\rho_i$ in a slightly different way, so as to also minimize the argument $Q_\text{jet}$. We introduce the minimal definition:
\begin{equation}
\rho_a^\text{minimal} = e^{-\hat{Y}} \qquad \qquad \rho_b^\text{minimal} = e^{\hat{Y}} \qquad \qquad \rho_J^\text{minimal} = \frac{1}{\cosh \eta}.
\label{eq:minimaldef}
\end{equation}
We can now perform a numerical study of the behavior of the cross section with respect to $\Tauc$ according to these three definitions, as shown in Figure~\ref{fig:taudef}.
We confirm what observed in~\cite{Campbell:2019gmd} for Higgs + 1 jet: the boosted definition reduces the impact of power corrections. The minimal definition has an even slightly better behavior for large values of $\Tauc$. In future studies, we plan to study the impact of power corrections for various new definitions of $\Tau$, corresponding to different combinations of the $\rho_i$ factors.
\section{Conclusions}
\label{sec:conc}

In this manuscript we have derived the 
next-to-leading power corrections to the $N$-jettiness factorization
theorem for 1-jet processes.  We have used the process of vector boson
plus jet as an illustrative example.  The NLP corrections can be written in a
simple analytic form and come from two sources: process-independent
phase space corrections, and process-dependent subleading power matrix
element corrections.  At the leading-logarithmic level the matrix element corrections can be
written in a universal form using results for next-to-leading
soft corrections, leading to a simple universal form for the NLP-LL
corrections.  At NLP the soft non-hemisphere terms contribute to the poles
and therefore give leading-logarithmic corrections to the cross section, unlike
at LP where they are finite.

We note that for the partonic process considered here as an example the
universal next-to-leading soft correction comes from gluon emission,
and is available in the literature.  It is known from color-singlet
production that soft quarks also contribute at the leading-logarithmic
level at NLP~\cite{Moult:2016fqy,Boughezal:2016zws}.  A corresponding form of the next-to-leading
soft corrections for quarks has yet to be derived.  We expect
that such an expression can be obtained.  Other possible future directions to expand upon this work include detailed numerical
studies of how different $\rho_i$ choices affect the size of the power
corrections, and the extension of this derivation to the NNLO level.

\section*{Acknowledgements}
 
R.~B. is supported by the DOE contract DE-AC02-06CH11357.  F.~P. is
supported by the DOE grants DE-FG02-91ER40684 and DE-AC02-06CH11357.
A.~I. is supported by the DOE grant DE-FG02-91ER40684 and the NSF
grant NSF-1520916.  This research used resources of the Argonne
Leadership Computing Facility, which is a DOE Office of Science User
Facility supported under Contract DE-AC02-06CH11357.

\appendix
\section{Phase space expansion coefficients}
\label{app:PS}

We compile here the expansion of the NLO phase space in $\Tau$ for the
various regions.

\subsection{Soft region}
\label{app:PSsoft}
We begin by studying the NLO parton momentum fractions.  In order to
determine the expansion coefficients in the two-jet case, we expand
the initial-state momentum fractions for small $\Tau_i$:
\begin{align}
\xi_a &= x_a+\frac{e^{Y+\eta} p_T \rho_a}{2 \sqrt{s} \sqrt{p_T^2+Q^2}}
\Tau_a + \left(1+\frac{e^{Y-\eta} p_T}{2 \sqrt{p_T^2 + Q^2}} \right)
\frac{\rho_b}{\sqrt{s}} \Tau_b -\frac{e^Y p_T \rho_J \cosh
  \eta}{\sqrt{s} \sqrt{p_T^2 + Q^2}} \Tau_J , \\
\xi_b &= x_b+\left(1+\frac{e^{-Y+\eta} p_T}{2 \sqrt{p_T^2 + Q^2}}
\right) \frac{\rho_a}{\sqrt{s}} \Tau_a +\frac{e^{-Y-\eta} p_T
  \rho_b}{2 \sqrt{s} \sqrt{p_T^2 + Q^2}} \Tau_b -\frac{e^{-Y} p_T
  \rho_J \cosh \eta}{\sqrt{s} \sqrt{p_T^2 + Q^2}} \Tau_J .
\end{align}
Using these expressions we can immediately derive the expansion coefficients for the
phase space measure of Eq.~(\ref{eq:PS2jetgeneral}):
\begin{align}
\Phi_\text{soft $2J$}^{(0,0,0)} &= f_q\left( x_a\right) f_{\bar{q}} \left( x_b\right), \\
\Phi_\text{soft $2J$}^{(1,0,0)}&=-f_q\left( x_a\right) f_{\bar{q}} \left( x_b\right) \frac{\rho_a}{ \sqrt{s} x_a x_b} \left[  x_a + \frac{e^{\eta} p_T }{2\sqrt{p_T^2 +Q^2}} \left( e^{-Y} x_a + e^Y x_b\right)\right]\nonumber \\
&+f'_q\left( x_a\right) f_{\bar{q}} \left( x_b\right)\frac{e^{Y+\eta} p_T \rho_a}{2\sqrt{s} \sqrt{p_T^2 +Q^2}} + f_q\left( x_a\right) f'_{\bar{q}}\left( x_b\right) \frac{\rho_a}{\sqrt{s}}\left(1+\frac{e^{-Y+\eta} p_T}{2\sqrt{p_T^2 + Q^2}} \right),\\
\Phi_\text{soft $2J$}^{(0,1,0)}&=-f_q\left( x_a\right) f_{\bar{q}} \left( x_b\right) \frac{\rho_b}{ \sqrt{s} x_a x_b} \left[  x_b + \frac{e^{-\eta} p_T }{2\sqrt{p_T^2 +Q^2}} \left( e^{-Y} x_a + e^Y x_b\right)\right]\nonumber \\
&+f'_q\left( x_a\right) f_{\bar{q}} \left( x_b\right) \frac{\rho_b}{\sqrt{s}}\left(1+\frac{e^{Y-\eta} p_T}{2\sqrt{p_T^2 + Q^2}} \right)
+ f_q\left( x_a\right) f'_{\bar{q}}\left( x_b\right)\frac{e^{-Y-\eta} p_T \rho_b}{2\sqrt{s} \sqrt{p_T^2 +Q^2}} ,\\
\Phi_\text{soft $2J$}^{(0,0,1)}&=f_q\left( x_a\right) f_{\bar{q}} \left( x_b\right)\frac{p_T \left( e^{-Y} x_a + e^Y x_b\right) \rho_J \cosh \eta}{\sqrt{s} \sqrt{p_T^2+Q^2} x_a x_b}\nonumber \\
& -f'_q\left( x_a\right) f_{\bar{q}}  \left( x_b\right)\frac{e^Y p_T \rho_J \cosh \eta}{\sqrt{s} \sqrt{p_T^2 +Q^2}}-f_q\left( x_a\right) f'_{\bar{q}}  \left( x_b\right)\frac{e^{-Y} p_T \rho_J \cosh \eta}{\sqrt{s} \sqrt{p_T^2 +Q^2}}.
\end{align}
The superscripts denote the orders in the $\Tau_a$, $\Tau_b$, and
$\Tau_J$ expansions of each term.  This matches the notation for the
soft-region expansion introduced in Eq.~(\ref{eq:Sfunctions}).

In the one-jet case, the expansion of the initial-state momentum fractions is
\begin{align}
\xi_a &= x_a +\frac{e^\eta \rho_J \cosh \eta}{\sqrt{s}} \Tau_J ,\\
\xi_b &= x_b+\frac{e^{-\eta} \rho_J \cosh \eta}{\sqrt{s}} \Tau_J .
\end{align}
The phase space expansion coefficients are
\begin{align}
\Phi_\text{soft $1J$}^{(0,0,0)} &= f_q\left( x_a\right) f_{\bar{q}}
\left( x_b\right), \\
\Phi_\text{soft $1J$}^{(1,0,0)}&=0,\\
\Phi_\text{soft $1J$}^{(0,1,0)}&=0 ,\\
\Phi_\text{soft $1J$}^{(0,0,1)}&=-f_q\left( x_a\right) f_{\bar{q}} \left( x_b\right) \frac{\rho_J \cosh \eta}{\sqrt{s}} \left(\frac{e^{-\eta}}{x_b} +\frac{e^\eta}{x_a}\right) \nonumber \\
&+f'_q\left( x_a\right) f_{\bar{q}} \left( x_b\right) \frac{e^\eta\rho_J \cosh \eta}{\sqrt{s}}+f_q\left( x_a\right) f'_{\bar{q}} \left( x_b\right) \frac{e^{-\eta} \rho_J \cosh \eta}{\sqrt{s}} .
\end{align}
To obtain our final form we must express $\Tau_J$ in terms of
$\Tau'_J=\Tau$, which we defined in Eq.~\eqref{eq:tauidef}:
\begin{equation}
\Tau_J = \Tau'_J \left( 1+\frac{\rho_J \Tau'_J \cosh \eta - e^\eta \rho_a \Tau_a - e^{-\eta} \rho_b \Tau_b}{2 p_T}\right).
\end{equation}
\subsection{Beam region}
\label{app:PSbeam}
In order to obtain the expansion coefficients of the beam-region phase
space, we first expand the initial-state momentum fractions for small $\Tau$:
\begin{align}
\xi_a = \frac{x_a}{z_a} + \frac{e^Y p_T \sqrt{x_a
  \rho_a}}{s^{1/4}\sqrt{p_T^2+Q^2}} \sqrt{\frac{1-z_a}{z_a}}\cos \phi
  \sqrt{\Tau} + \frac{e^Y x_a \rho_a \left(Q^2 + p_T^2 \sin^2 \phi
  \right)}{2 \left(p_T^2 +Q^2 \right)^{3/2}} \left(\frac{1-z_a}{z_a}
  \right) \Tau ,
\label{eq:xiabeam}
\end{align}
\begin{align}
\xi_b = x_b +\frac{e^{-Y} p_T \sqrt{x_a \rho_a}}{s^{1/4}\sqrt{p_T^2+Q^2}} \sqrt{\frac{1-z_a}{z_a}}\cos \phi \sqrt{\Tau} +\left[ \frac{e^Y x_a \rho_a \left(Q^2 + p_T^2 \sin^2 \phi \right)}{2 \left(p_T^2 +Q^2 \right)^{3/2}} \left(\frac{1-z_a}{z_a} \right)+\frac{\rho_a }{\sqrt{s}}\right] \Tau.
\label{eq:xibbeam}
\end{align}
Upon substitution in Eq.~(\ref{eq:PS2jetgeneral}) these lead to the
following phase space coefficients, relevant at LL:
\begin{align}
\Phi_\text{beam $a$}^{\left(0,0\right)}&=\Phi_\text{beam $b$}^{\left(0,0\right)}= f_q\left( x_a\right) f_{\bar{q}},\\
\Phi_\text{beam $a$}^{\left(1/2,1/2\right)}(\phi) &=\frac{p_T \cos \phi}{s^{1/4}}\sqrt{\frac{ \rho_a x_a}{p_T^2 + Q^2}} \Bigg[-  f_q \left(x_a  \right) f_{\bar{q}} \left( x_b\right) \left( \frac{e^{-Y}}{x_b}+\frac{e^{Y} }{x_a}\right)\nonumber \\
&+e^Y f'_q \left(x_a \right) f_{\bar{q}} \left( x_b\right)+e^{-Y} f_q \left(x_a\right) f'_{\bar{q}} \left( x_b\right) \Bigg] ,\\
\Phi_\text{beam $a$}^{\left(1,0\right)} &= \frac{\rho_a}{\sqrt{s} x_b} \left[-f_q \left(x_a\right) f_{\bar{q}} \left( x_b\right) + x_b f_q \left(x_a\right)f'_{\bar{q}} \left( x_b\right) \right],\\
\Phi_\text{beam $a$}^{\left(0,1\right)} &= -f_q \left(x_a\right) f_{\bar{q}} \left( x_b\right) + x_a f'_q \left(x_a\right)f_{\bar{q}} \left( x_b\right) ,\\
\Phi_\text{beam $b$}^{\left(1/2,1/2\right)}(\phi) &=\frac{p_T \cos \phi}{s^{1/4}}\sqrt{\frac{ \rho_b x_b}{p_T^2 + Q^2}} \Bigg[-  f_q \left(x_a  \right) f_{\bar{q}} \left( x_b\right) \left( \frac{e^{-Y}}{x_b}+\frac{e^{Y} }{x_a}\right)\nonumber \\
&+e^Y f'_q \left(x_a \right) f_{\bar{q}} \left( x_b\right)+e^{-Y} f_q \left(x_a\right) f'_{\bar{q}} \left( x_b\right) \Bigg] ,\\
\Phi_\text{beam $b$}^{\left(1,0\right)} &= \frac{\rho_b}{\sqrt{s} x_a} \left[-f_q \left(x_a\right) f_{\bar{q}} \left( x_b\right) + x_a f'_q \left(x_a\right)f_{\bar{q}} \left( x_b\right) \right],\\
\Phi_\text{beam $b$}^{\left(0,1\right)} &= -f_q \left(x_a\right) f_{\bar{q}} \left( x_b\right) + x_b f_q \left(x_a\right)f'_{\bar{q}} \left( x_b\right) .
\end{align}
%
\subsection{Jet region}
\label{app:PSjet}
The first step in determining the expansion coefficients in the jet
region is to expand $\Tau_J$ in terms of $\Tau=\Tau'_J$ :
\begin{equation}
\Tau_J=\left( 1-z_J\right) \Tau +\sqrt{\frac{2 \cosh\eta \rho_J}{p_T}} \sqrt{(1-z_J)z_J} \cos \phi \Tau^{3/2} +\frac{\rho_J \cosh\eta}{2 p_T} \left( -1+3 z_J + z_J \cos(2 \phi)\right) \Tau^2.
\end{equation}
We then expand the transverse mass of the jet:
\begin{equation}
m_T = p_T + \rho_J \cosh \eta \Tau.
\end{equation}
Finally, we can derive the phase space expansion coefficients upon
substituting these expressions into Eq.~(\ref{eq:PS1jetgeneral}):
\begin{align}
\Phi_\text{jet}^{(0,0)}&=f_q\left( x_a\right) f_{\bar{q}} \left(x_b \right),\\
\Phi_\text{jet}^{(1/2,1/2)}(\phi)&= \frac12 \sqrt{\frac{2 \rho_J \cosh \eta}{p_T}} f_q\left( x_a\right) f_{\bar{q}} \left(x_b \right) \cos \phi,\\
\Phi_\text{jet}^{(1,0)}&=\frac{ \rho_J \cosh \eta}{\sqrt{s} } \left[-\left(\frac{e^{\eta} }{x_a}+\frac{e^{-\eta}}{x_b}\right) f_q\left(x_a\right) f_{\bar{q}}\left(x_b\right) + e^{\eta}  f'_q\left(x_a\right) f_{\bar{q}}\left(x_b\right)+e^{-\eta} f_q\left(x_a\right) f'_{\bar{q}} \left(x_b\right)\right],\\
\Phi_\text{jet}^{(0,1)}&=0.
\end{align}
%


\end{document}